\documentclass[11pt]{article}
\usepackage{amsmath,amsfonts,amssymb}
\usepackage{graphicx}
\usepackage[round,sort&compress]{natbib}
\usepackage[letterpaper=true,colorlinks=true,pdfpagemode=none,urlcolor=blue,linkcolor=blue,citecolor=blue,pdfstartview=FitH]{hyperref}
\usepackage{xcolor}
\usepackage{booktabs}
\usepackage{enumitem}

\usepackage[margin=1in]{geometry}

\usepackage{algorithm}
\usepackage{algpseudocode}
\usepackage{makecell}

\newcommand{\E}{\mathbb{E}}
\newcommand{\email}[1]{\href{mailto:#1}{#1}}

\DeclareMathOperator*{\argmax}{argmax}

\renewcommand{\hat}{\widehat}

\newcommand{\yobs}[1]{w_{#1}}
\newcommand{\cobs}[1]{\pi_{#1}}
\newcommand{\iobs}[1]{i_{#1}}
\newcommand{\wobs}[1]{c_{#1}}
\newcommand{\lobs}[1]{l_{#1}}
\newcommand{\robs}[1]{r_{#1}}
\newcommand{\piobs}[1]{y_{#1}}

\newcommand{\youtput}{wages}
\newcommand{\Output}{Wages}
\newcommand{\consumption}{inflation}

\newcommand{\investment}{investment}

\newcommand{\hours}{hours worked}
\newcommand{\realinterest}{real interest}
\newcommand{\labor}{labor}
\newcommand{\interest}{interest}
\newcommand{\wage}{consumption}
\newcommand{\wages}{consumption}
\newcommand{\price}{price}
\newcommand{\prices}{prices}
\newcommand{\inflation}{output}
\newcommand{\Inflation}{Output}
\newcommand{\inflationrate}{output}
\newcommand{\working}{working}
\newcommand{\consuming}{inflating}
\newcommand{\leisure}{leisure}
\newcommand{\spending}{\consuming}

\usepackage{booktabs}
\usepackage{longtable}
\usepackage{array}
\usepackage{multirow}
\usepackage{wrapfig}
\usepackage{float}
\usepackage{colortbl}
\usepackage{pdflscape}
\usepackage{tabu}
\usepackage{threeparttable}
\usepackage{threeparttablex}
\usepackage[normalem]{ulem}
\usepackage{makecell}
\usepackage{xcolor}

\begin{document}


\title{Empirical Macroeconomics and DSGE Modeling in Statistical
Perspective}
\author{  
    Daniel J. McDonald\thanks{Both authors were supported by a grant
(INO1400020) from the Institute for New Economic Thinking. We are
grateful for advice and comments to David N. DeJong, J. Bradford DeLong
and Robert Waldmann (none of whom should be held responsible for our
choices). DJM was partially supported by the National Science Foundation
(grants DMS1407439 and DMS1753171) and the National Sciences and
Engineering Research Council of Canada (RGPIN-2021-02618). CRS wishes to
acknowledge support from the National Science Foundation (grants
DMS1207759 and DMS1418124), and valuable conversations over many years
with Zmarak M. Shalizi.}  \\
  Department of Statistics\\
University of British Columbia\\
Vancouver, BC Canada\\
\email{daniel@stat.ubc.ca}  
   \and   Cosma Rohilla Shalizi \\
  Departments of Statistics and Data Science and of Machine Learning\\
Carnegie Mellon University\\
Pittsburgh, PA USA\\
\emph{and} Santa Fe Institute\\
Santa Fe, NM USA\\
\email{cshalizi@cmu.edu}  
  }
\date{}

\maketitle

\begin{abstract}
  Dynamic stochastic general equilibrium (DSGE) models have been an
ubiquitous, and controversial, part of macroeconomics for decades. In
this paper, we approach DSGEs purely as statstical models. We do this by
applying two common model validation checks to the canonical
\citet{SmetsWouters2007} DSGE: (1) we simulate the model and see how
well it can be estimated from its own simulation output, and (2) we see
how well it can seem to fit nonsense data. We find that (1) even with
centuries' worth of data, the model remains poorly estimated, and (2)
when we swap series at random, so that (e.g.) what the model gets as the
inflation rate is really hours worked, what it gets as hours worked is
really investment, etc., the fit is often only slightly impaired, and in
a large percentage of cases actually \emph{improves} (even out of
sample). Taken together, these findings cast serious doubt on the
meaningfulness of parameter estimates for this DSGE, and on whether this
specification represents anything structural about the economy.
Constructively, our approaches can be used for model validation by
anyone working with macroeconomic time series. 
  
  \vspace{11pt}
\end{abstract}

\hypertarget{introduction}{%
\section{Introduction}\label{introduction}}

Since the 1980s, academic macroeconomics has been dominated by dynamic
stochastic general equilibrium (DSGE) models, and economists have
devoted a much attention to their specification, elaboration,
mathematical manipulation, estimation, and theoretical refinement. This
rise to prominence has been opposed, and there have always been critics
of the whole approach on theoretical grounds, charging that, in one way
or another, DSGE models are (or embody, or presume) bad economic
theories. Instead of questioning whether DSGEs are good
\emph{economics}, we look at whether they are good \emph{models}, i.e.,
whether they meet common, intuitive standards of statistical modeling.

In this paper, we are not primarily concerned with whether DSGE models
can predict macroeconomic variables outside of the time period used to
estimate their parameters. Out-of-sample forecasting is a quite
elementary test of any statistical model's actual fit to the data
\citep{tEoSL-2nd}, and this has been appreciated for a very long time
\citep{Stone1974,Geisser-predictive-sample-reuse,Geisser-Eddy-predictive-approach}.
Moreover, it's well-established that DSGEs are bad at it
\citep{Edge-Gurkaynak-on-dsges}, and even vector autoregressions do much
better. We thus take it for granted that DSGEs are \emph{currently} no
good for prediction, and ask whether they will \emph{ever} be capable of
accurate forecasts. By the simple, standard device of simulating a DSGE
and then fitting the same model to the simulation output, we show
(sec.~\ref{sec:simulate-estimate} below) that even when correctly
specified in their entirety and provided with centuries of simulated
data, these models remain incapable of forecasting. Worse, their
parameters remain very badly estimated. To reliably estimate models of
such coelenterate flexibility would require thousands of years (at
least) of data from a stationary economy.

We are aware that many economists downplay using DSGEs for prediction on
the grounds that the models are instead supposed to inform us about the
structure of the economy and about the consequences of prospective
policy interventions. Even if this is true, however, one would need some
reason to think that \emph{this} DSGE, rather than another, was getting
things right. While we agree that models which are capable of accurate
statistical prediction may be horrible at causal, counter-factual
prediction, it has long been understood that the reverse is not true
\citep{Spirtes-Glymour-Scheines-1st, Pearl-causality}, and this is now
literally a textbook point in causal inference
\citep{Morgan-Winship-counterfactuals-2nd}. A model which gets the
causal structure right, and can make accurate counter-factual
predictions, should \emph{a fortiori} be capable of accurate statistical
prediction as well. Hence economists' confidence in their favorite DSGEs
suitability for policy evaluation cannot be rooted in their statistical
predictive ability, since the later does not exist.

We also undermine the notion that DSGE specifications capture the
structure of the economy by the simple expedient of swapping the
different time series on which they are fit --- giving the model as
``investment'' the series that is really ``hours worked'', and so forth
(sec.~\ref{sec:permutation-summary} below). Not only does such series
swapping do little to degrade the DSGE's performance, in sample or out
of sample, in a large fraction of permutations it actually
\emph{enhances} predictability. It is, of course, open to an economist
to maintain their belief in a favorite DSGE's capturing the structure of
the macroeconomy even if it cannot predict, and cannot tell the
difference between the real data and one with all the series swapped,
but such faith truly is maintained on the evidence of things not seen.

The principles to which we appeal --- using simulation to assess
estimation methods; using randomization to gauge how well a model can
appear to fit nonsense data --- are not recondite points of mathematical
theory, and involve no controversial questions in the foundations of
statistics. Rather, they are readily explained notions, accepted on
pretty much all sides within modern statistics, whose force is easily
grasped once they are presented. We return to their implications for
macroeconomic modeling in the conclusion.

\hypertarget{brief-remarks-on-the-literature}{%
\subsection{Brief Remarks on the
Literature}\label{brief-remarks-on-the-literature}}

There is a familiar story to the rise of DSGEs from the 1970s to the
1990s, which roots them in, on the one hand, the desire to give
``microfoundations'' to macroeconomic models, and, on the other hand,
the \citet{Lucas1976} critique of ``old Keynesian'' models, and the
latter's apparent empirical failure in the 1970s. The breakthrough
paper, on this account, was the ``real business cycle'' model of
\citet{KydlandPrescott1982}. Versions of this story are familiar from
textbooks (e.g., \citealt{DeJongDave2007}), and it's not our place here
to dispute it. It is undeniably true that, following
\citet{KydlandPrescott1982}, macroeconomists rapidly cultivated many
breeds of DSGE model, embodying a range of substantive economic
assumptions, but also conserving certain crucial features that define
the lineage, such as the use of representative agents (often just one
representative agent) in equilibrium. The motivation was also conserved:
DSGEs (it is held) clearly separate enduring, ``structural'' aspects of
the economy, ultimately relating to tastes, technologies and
institutions, from policy and from fluctuations. The structural,
tastes-and-technologies aspects are held to be invariant, at least to
policy and at least over the relevant time scales. Embedding these into
a dynamic equilibrium model is supposed to get around the Lucas
critique, by ensuring that even if the predictions of the model are
known and are used as the basis of policy, the model will remain valid,
\emph{because} it describes an equilibrium over time.

The model we focus on, that of \citet{SmetsWouters2007}, hereafter
denoted SW, while now some years old, is still widely regarded as an
acceptable baseline model. A great deal of macroeconomic theorizing thus
consists of taking this model, or other very similar ones, and
elaborating on them by adding new frictions or shocks. These additions
are generally intended to accommodate observed phenomena, to incorporate
conjectured mechanisms, or both. It is not our purpose here to comment
on specific descendants of the SW or other baseline DSGE models. We
merely note that the strategy of responding to issues or defects with
the model by always making it \emph{more} complex (more frictions, more
shocks, etc.) is at least questionable from a statistical perspective.

As we wrote in the introduction, our objective here is not to add to the
literature for or against DSGE models from the viewpoint of economic
theory. It is not even pertinent to survey that voluminous, often
acrimonious, literature. We take no sides, here, on whether agents in
macroeconomic models should be assumed to have rational (rather than
adaptive) expectations, should have perfect (rather than bounded)
rationality, should (or should not) obey the Euler equations, etc. We do
not even take a side on whether aggregating large numbers of
heterogeneous economic actors into a few representative agents is in
fact ``microfounded''\footnote{But on this point, we cannot resist
  noting that it is hard to see how to get around the difficulties
  raised by \citet{Kirman-contra-representative-agent} and, more
  especially,
  \citet{Jackson-Yariv-non-existence-of-representative-agents}.}. Those
who are \emph{a priori} inclined to dismiss DSGEs as inappropriate
models will, we suspect, find our results on their statistical
difficulties congenial\footnote{Except, perhaps, for those who also
  object to quantitative empirical economics in the first place.}. But
no result we are aware of implies that theoretically sound economic
models must also have good statistical properties. Moreover, for all we
know at this stage, all the proposed alternatives to DSGEs \emph{also}
suffer from the same statistical flaws! (We invite their partisans, or
perhaps their enemies, to check.)

Turning to the literature on statistical properties of DSGEs, this has
mostly focused on issues of estimation and testing; we cite specific
works below as needed. A smaller body of work has examined out-of-sample
forecasting ability of DSGEs, notably \citet{Edge-Gurkaynak-on-dsges}
(see also references therein). That paper found that these model
predicted poorly, but ingeniously attributed this to the close control
exercises by central bankers and other policy-makers. (Other
explanations suggest themselves.) Another tangentially-relevant body of
work has examined identification issues in DSGEs. The most important
paper here is probably still
\citet{Komunjer-Ng-identification-of-DSGEs}, which gave fairly practical
algebraic conditions for checking the identifiability of DSGEs. (This
built off earlier work by \citealt{Iskrev2009} among others.)
Identification means that all model parameters can (in principle) be
estimated in the limit; estimates using the available, non-asymptotic
amount of data might be unacceptably imprecise, and might converge
unacceptably slowly. We are unaware of any previous work which has
addressed the practical estimability of DSGEs by the means used here, or
anything like it. We are also unaware of any previous work which probes
whether DSGEs are actually capturing the structure of the macroeconomy
by anything like the expedient of series swapping.

\clearpage

\hypertarget{specification-and-baseline-estimation-of-the-dsge}{%
\section{Specification and Baseline Estimation of the
DSGE}\label{specification-and-baseline-estimation-of-the-dsge}}

In this section, we review how DSGE models work, the specification of
the \citet{SmetsWouters2007} DSGE that serves as our test case, and some
of the issues that arise in estimating the parameters of this model.

\hypertarget{sec:solveDSGEs}{%
\subsection{Solving DSGEs}\label{sec:solveDSGEs}}

A DSGE model is the solution to a constrained stochastic inter-temporal
optimization problem \begin{align}
  z_t^* &=\argmax \sum_{t=0}^\infty \E g(z_t) &&\mbox{s.t.}&z_t&=h(z_{t-1}),
\end{align} for some nonlinear functions \(g\) and \(h\) parametrized by
a \(k\)-dimensional vector of ``deep'' parameters \(\theta\). The
economic agents posited by the model are assumed to solve this problem
(optimally) at each time \(t\) conditional on all current and previously
available information, and we observe part of the solution: that is the
observable data are \(x_t\) which is a subset of the indices of
\(z_t^*\).

To estimate a DSGE, the first step is to solve the optimization problem
by deriving the first-order conditions for an optimum. The resulting
nonlinear system can be written as \begin{equation}
  \Phi(z_t,\ z_{t+1})=0,
\end{equation} for some \(n\)-dimensional function \(\Phi\). Such
systems can rarely be solved analytically for the optimal path; instead,
the first-order conditions are used to express the model in terms of
stationary variables, and the system is linearized around this
steady-state (that is, let \(z\) be the vector satisfying
\(z=z_t=z_{t-1}\)). Upon writing \(z_t = z^*_t - z\), the (linearized)
DSGE can be written as \begin{equation}
  \Gamma_0 z_t = \Gamma_1 z_{t-1} + A + B \epsilon_t + C \eta_t
  \label{eq:lin-dsge}
\end{equation} in the notation of \citet{Sims2002}. Here \(\epsilon_t\)
are exogenous, possibly serially correlated, random disturbances and
\(\eta_t=z_t - \E_t z_{t+1}\) are expectational errors determined as
part of the model solution. The matrices \(\Gamma_0\), \(\Gamma_1\),
\(A\), \(B\), and \(C\) are functions of \(\theta\).

There are a number of ways to estimate \(\theta\) using data; a complete
treatment is beyond our scope here, but see \citet{DeJongDave2007} for
details. We will focus on likelihood-based approaches, and so we must
solve the model in equation \eqref{eq:lin-dsge}. There are many
approaches to solving linear rational expectations models
\citep[e.g.][]{BlanchardKahn1980,Klein2000}, but we will use that in
\citet{Sims2002} due to its ubiquity. This method essentially uses a QZ
factorization to remove \(\Gamma_0\) from the left side of the equation
while correctly handling explosive components of the model (merely
multiplying through by the generalized inverse of \(\Gamma_0\),
\(\Gamma_0^\dagger\), can lead to portions of the product
\(\Gamma_0^\dagger \Gamma_1\) implying nonstationarity). Following this
procedure, we retrieve a system of the form \begin{equation}
  \label{eq:ss-system}
  z_t = d + T z_{t-1} + H \varsigma_t
\end{equation} as long as there is a unique mapping from equation
\eqref{eq:lin-dsge} (there may be multiple solutions or none depending
on \(\theta\)). Since some of the \(z_t\) are unobserved, we augment the
transition equation in \eqref{eq:ss-system} with an observation equation
\begin{equation}
  \label{eq:obs-equation}
  x_t = Z z_t,
\end{equation} where the matrix \(Z\) subselects the appropriate
elements of \(z_t\). Collecting equations \eqref{eq:ss-system} and
\eqref{eq:obs-equation} gives the form of a linear state-space model.
Assuming that the errors \(\varsigma_t\) are serially independent
multivariate Gaussian allows us to evaluate the likelihood of some
parameter vector \(\theta\) given observed data \(x_1,\ldots,x_T\).
Evaluating the likelihood can be done using the Kalman filter
\citep{Kalman1960} which is readily available in most software packages.
The procedure outlined above is summarized in \autoref{alg:solveRE}.

\begin{algorithm}[t]
\begin{algorithmic}[1]
  \State Write down DSGE as a constrained optimization problem.
  \State Determine the first order conditions for an optima.
  \State Determine steady state path of observables.
  \State Linearize the DSGE around the steady-state yielding a general
    form for \eqref{eq:lin-dsge}.
  \Procedure{Estimate the model}{$\theta$}
    \State Fix a plausible parameter vector $\theta$.
    \State For this $\theta$, cast \eqref{eq:lin-dsge} into state-space form
      using the method of \citet{Sims2002}.
    \State Using equations \eqref{eq:ss-system} and \eqref{eq:obs-equation}, evaluate the
      likelihood of $\theta$ using the Kalman filter.
    \State Maximize the likelihood or explore the posterior, repeating
      as needed at new values of $\theta$.
    \EndProcedure
  \end{algorithmic}
  \caption{Pseudoalgorithm for estimating linear rational expectations models}
  \label{alg:solveRE}
\end{algorithm}

\subsection{Estimating the \citet{SmetsWouters2007} model}
\label{sec:estim-sw-model}

In this section, we provide a description of the procedure we use for
estimating the \citet{SmetsWouters2007} model. The model is now standard
in the macroeconomic forecasting literature, and, though code is readily
available (for example in \texttt{Dynare} or
\texttt{Matlab}\footnote{See also
  \url{https://www.aeaweb.org/articles.php?doi=10.1257/aer.97.3.586&fnd=s}}),
our version is implemented fully in \texttt{R}. A complete description
of the economic implications of the model and its log-linearized form
can be found in \citet{SmetsWouters2007} as well as \citet{Iskrev2009}
and other sources.

For this model, we use seven observable data series: output growth,
consumption growth, investment growth, real wage growth, inflation,
hours worked, and the nominal interest rate, which we collect into the
vector \(x_t\) \begin{equation}
  x_t = \begin{bmatrix} y_t-y_{t-1} & c_t - c_{t-1} & i_t-i_{t-1} &
    w_t-w_{t-1} & \pi_t & l_t & r_t \end{bmatrix}^\top.
\end{equation} We describe the specific data and preprocessing routines
in \autoref{sec:data-preprocessing}, but we note that all the data are
publicly available from the Federal Reserve Economic Database
(\href{http://research.stlouisfed.org/fred2/}{FRED}). We use data from
the first quarter of 1956 until the fourth quarter of 2018. Following
preprocessing, we are left with 251 available time points.

The model has 52 ``deep'' parameters as well as 7 parameters
representing the standard deviations of the stochastic shocks. Of these
59 total parameters, we estimate 36: 18 are derived from steady-state
values as functions of other parameters and 5 are fixed a priori (as in
\citealt{SmetsWouters2007}). The prior distributions for the 36
estimated parameters are given in
\autoref{tab:parameter-estimate-table}. To estimate the model we
minimize the negative log likelihood, penalized by the prior. This is
the same as finding the maximum \emph{a posteriori} estimate in a
Bayesian setting. Because the likelihood is ill-behaved, having many
flat sections as well as local minima, we used \texttt{R}'s
\texttt{optimr} package. We estimated the parameter using both the
simulated annealing method, which stochastically explores the likelihood
surface in a principled manner, and the conjugate gradient technique.
Each procedure was started at 5 random initializations (drawn from the
prior distribution) and run for 50,000 iterations (likelihood
evaluations) for each starting point. We train the model using only the
first 200 time points, saving the remainder to evaluate the model's
(pseudo-out-of-sample) predictive performance.

\autoref{tab:parameter-estimate-table} presents the posterior mode based
on our procedure. Note first that some of the parameter estimates are
similar to those presented in \citet{SmetsWouters2007} (shown in the
last column), while others differ dramatically. However, comparing the
likelihood of of our estimated parameters to those in
\citet{SmetsWouters2007}, our fit is significantly better. For our
dataset, the penalized negative log likelihood of the parameters is 1145
compared to 1232, an improvement of more than 7.1\%. The result is
similar for the unpenalized negative log likelihood. To check for
robustness, we also ran the optimization procedure for 1 million
parameter draws and the likelihood only decreased by about 1\%. We are
therefore confident that we have a parameter combination which nearly
achieves the global optimum.\footnote{We have been unable to explain why
  \citet{SmetsWouters2007} gives such a different estimate of the
  posterior mode. It is worth pointing out however, that their MCMC is
  not necessarily attempting to find the maximum of the posterior,
  although it may locate one.}

\begin{table}

\caption{\label{tab:parameter-estimate-table}Prior distributions, posterior modes, and posterior mode as
      estimated by \citet{SmetsWouters2007} for the 36 estimated
      parameters. All values are rounded to two decimal places.}
\centering
\fontsize{9}{11}\selectfont
\begin{tabular}[t]{cccccccc}
\toprule
 & \makecell[c]{prior\\distribution} & \makecell[c]{prior\\mean} & \makecell[c]{prior\\stdev} & \makecell[c]{lower\\bound} & \makecell[c]{upper\\bound} & \makecell[c]{posterior\\mode} & \makecell[c]{SW posterior\\mode}\\
\midrule
$\sigma_a$ & igamma & 0.10 & 2.00 & 0.01 & 3.00 & 0.44 & 0.45\\
$\sigma_b$ & igamma & 0.10 & 2.00 & 0.03 & 5.00 & 0.30 & 0.24\\
$\sigma_g$ & igamma & 0.10 & 2.00 & 0.01 & 3.00 & 0.57 & 0.52\\
$\sigma_I$ & igamma & 0.10 & 2.00 & 0.01 & 3.00 & 0.48 & 0.45\\
$\sigma_r$ & igamma & 0.10 & 2.00 & 0.01 & 3.00 & 0.22 & 0.24\\
\addlinespace
$\sigma_p$ & igamma & 0.10 & 2.00 & 0.01 & 3.00 & 0.15 & 0.14\\
$\sigma_w$ & igamma & 0.10 & 2.00 & 0.01 & 3.00 & 0.27 & 0.24\\
$\rho_a$ & beta & 0.50 & 0.20 & 0.01 & 1.00 & 0.97 & 0.95\\
$\rho_b$ & beta & 0.50 & 0.20 & 0.01 & 1.00 & 0.14 & 0.18\\
$\rho_g$ & beta & 0.50 & 0.20 & 0.01 & 1.00 & 0.95 & 0.97\\
\addlinespace
$\rho_I$ & beta & 0.50 & 0.20 & 0.01 & 1.00 & 0.66 & 0.71\\
$\rho_r$ & beta & 0.50 & 0.20 & 0.01 & 1.00 & 0.13 & 0.12\\
$\rho_p$ & beta & 0.50 & 0.20 & 0.01 & 1.00 & 0.99 & 0.90\\
$\rho_w$ & beta & 0.50 & 0.20 & 0.00 & 1.00 & 0.95 & 0.97\\
$\mu_p$ & beta & 0.50 & 0.20 & 0.01 & 1.00 & 0.88 & 0.74\\
\addlinespace
$\mu_w$ & beta & 0.50 & 0.20 & 0.01 & 1.00 & 0.90 & 0.88\\
$\phi_1$ & gaussian & 4.00 & 1.50 & 2.00 & 15.00 & 5.73 & 5.48\\
$\sigma_c$ & gaussian & 1.50 & 0.38 & 0.25 & 3.00 & 1.55 & 1.39\\
$h$ & beta & 0.70 & 0.10 & 0.00 & 0.99 & 0.72 & 0.71\\
$\xi_w$ & beta & 0.50 & 0.10 & 0.30 & 0.95 & 0.78 & 0.73\\
\addlinespace
$\sigma_l$ & gaussian & 2.00 & 0.75 & 0.25 & 10.00 & 2.01 & 1.92\\
$\xi_p$ & beta & 0.50 & 0.10 & 0.50 & 0.95 & 0.60 & 0.65\\
$\iota_w$ & beta & 0.50 & 0.15 & 0.01 & 0.99 & 0.34 & 0.59\\
$\iota_p$ & beta & 0.50 & 0.15 & 0.01 & 0.99 & 0.26 & 0.22\\
$\Psi$ & beta & 0.50 & 0.15 & 0.01 & 1.00 & 0.57 & 0.54\\
\addlinespace
$\Phi$ & gaussian & 1.25 & 0.12 & 1.00 & 3.00 & 1.13 & 1.61\\
$r_\pi$ & gaussian & 1.50 & 0.25 & 1.00 & 3.00 & 2.06 & 2.03\\
$\rho$ & beta & 0.75 & 0.10 & 0.50 & 0.98 & 0.83 & 0.81\\
$r_y$ & gaussian & 0.12 & 0.05 & 0.00 & 0.50 & 0.10 & 0.08\\
$r_{\Delta y}$ & gaussian & 0.12 & 0.05 & 0.00 & 0.50 & 0.21 & 0.22\\
\addlinespace
$\overline{\pi}$ & gamma & 0.62 & 0.10 & 0.10 & 2.00 & 0.63 & 0.81\\
$100(\beta^{-1} -1)$ & gamma & 0.25 & 0.10 & 0.01 & 2.00 & 0.13 & 0.16\\
$\overline{l}$ & gaussian & 0.00 & 2.00 & -10.00 & 10.00 & 3.34 & -0.10\\
$\overline{\gamma}$ & gaussian & 0.40 & 0.10 & 0.10 & 0.80 & 0.47 & 0.43\\
$\rho_{ga}$ & gaussian & 0.50 & 0.25 & 0.01 & 2.00 & 0.63 & 0.52\\
\addlinespace
$\alpha$ & gaussian & 0.30 & 0.05 & 0.01 & 1.00 & 0.28 & 0.19\\
\bottomrule
\end{tabular}
\end{table}

\clearpage

\hypertarget{sec:simulate-estimate}{%
\section{Simulate and estimate}\label{sec:simulate-estimate}}

A simple way to evaluate a stochastic model is to simulate a long time
series from the model and see how well we can estimate the generating
process given more and more data. In a sense, this is a minimal
requirement for any model: does it produce consistent estimates of the
true parameters? And furthermore, how much data will we need?
Statistical theory for independent and identically distributed data says
that maximum likelihood estimators for a fixed number of parameters are
\(n\)-consistent in MSE. This is essentially as fast as we could hope.
In this exercise, we abstract from the non-linear DSGE: supposing that
the data were actually generated by the linearized DSGE represented by
equations \eqref{eq:ss-system} and \eqref{eq:obs-equation}, can we
recover the ``deep'' parameters which generated the data?

To answer this (relatively simple) question, we generate data using the
parameters estimated above and presented in
\autoref{tab:parameter-estimate-table}. To reduce dependence on initial
conditions, we simulate 3100 data points and discard the first 1000. We
then train the model using the first 100 data points (25 years worth of
quarterly data), and try to predict the next 1000 to measure
out-of-sample performance. We then increment forward 20 time-steps (5
years), train again and predict the next 1000 observations. We continue
this process until we are using 1100 data points (275 years of data) to
estimate the model. To average over simulation error, we repeat the
entire exercise 100 times. For each estimation, we initialize the
optimization procedure at the true parameter value and run it for 30000
iterations. This ensures that the true parameters are considered,
maximizing the chance that the estimation will return the data
generating values. Our goal here is to learn whether we could hope to
learn the right parameters in 275 years' time if the linearized DSGE is
actually true, and furthermore, how well can we predict future economic
movements under this idealized scenario.\footnote{It is, of course,
  wildly implausible to imagine a 275 year interval which does
  \emph{not} see massive changes to tastes, technologies and
  institutions. The differences between 1746 and 2021 are obvious, but
  your favorite economic historian will explain at length that these
  were all transformed between 1471 and 1746, too. Recognizing this
  hardly makes DSGEs \emph{more} attractive.}

All figures in this section show the mean in red along with 30\%, 60\%,
80\%, and 90\% confidence bands calculated from the replications.

\begin{figure}[t]

{\centering \includegraphics{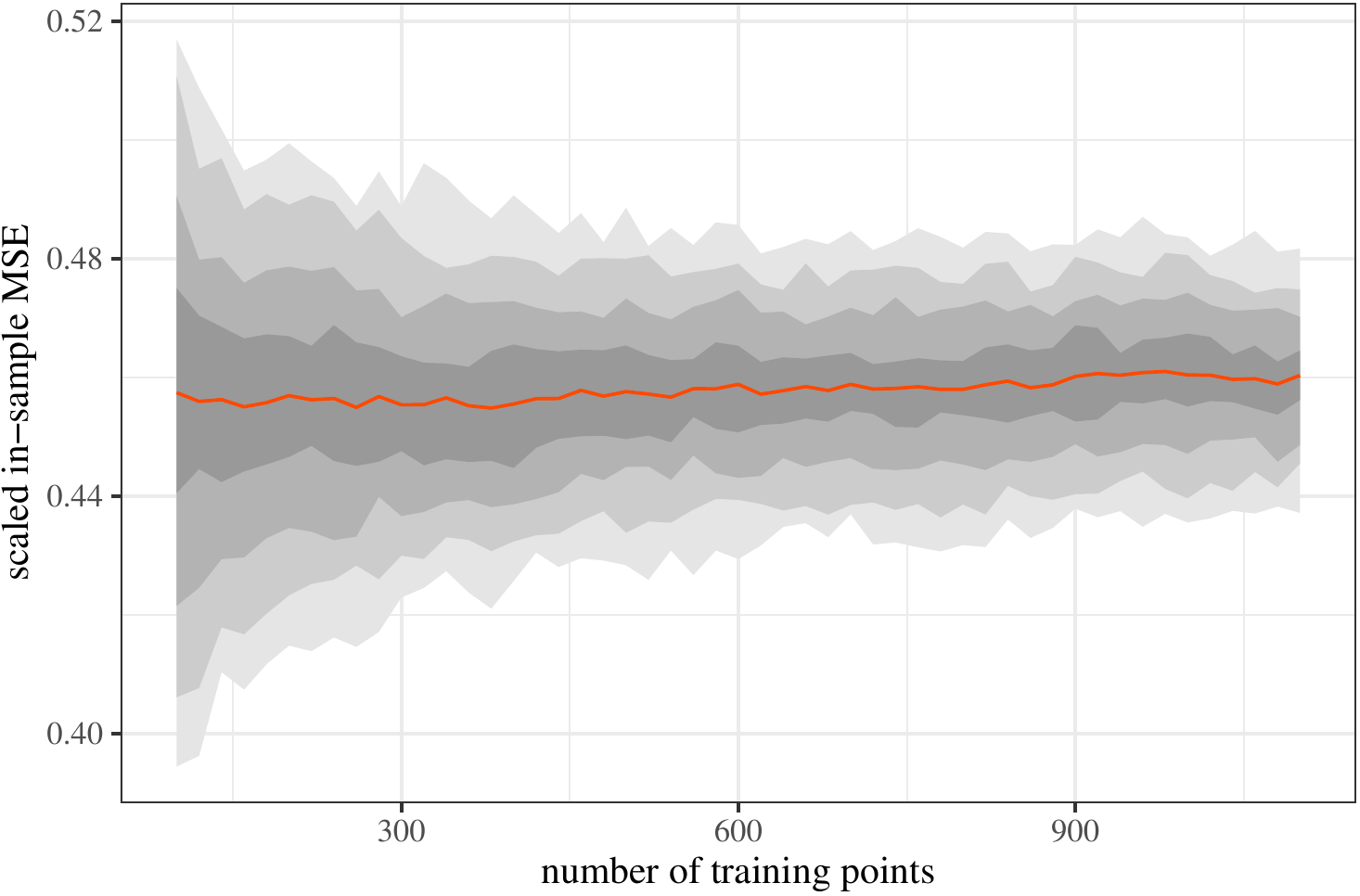} 

}

\caption{Training (in sample) error averaged across series as the number of training points increases. We would expect the variability to decline while the average remains constant. The mean is shown in red along with 30\%, 60\%, 80\%, and 90\% confidence bands calculated from the replications.}\label{fig:train-error}
\end{figure}

\begin{figure}[t]

{\centering \includegraphics{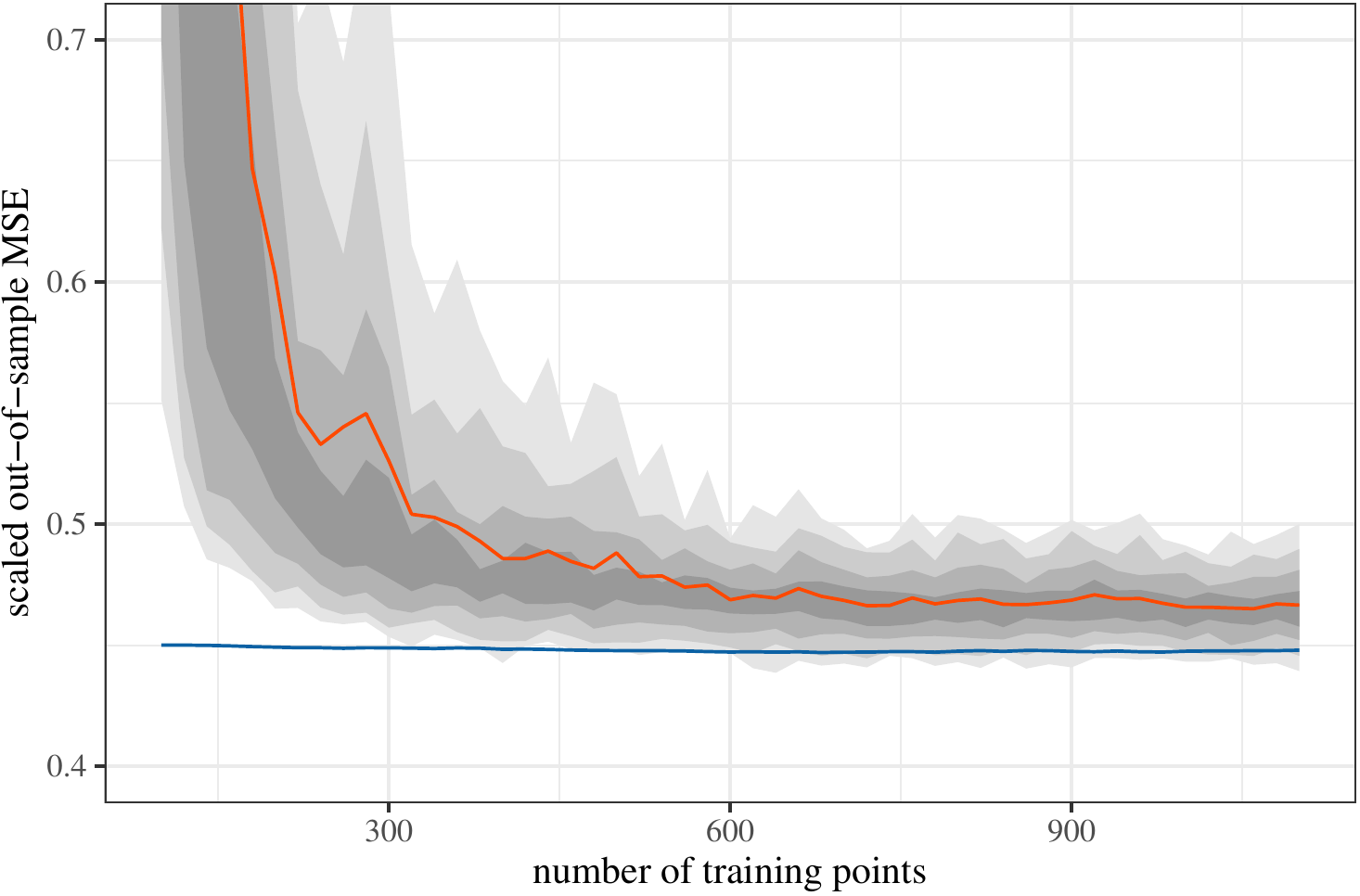} 

}

\caption{Out-of-sample error averaged across series as the number of training points increases. We would expect both the variability and the average to decline. The blue line is the test error for the true model.}\label{fig:test-error}
\end{figure}

\autoref{fig:train-error} shows the average training error (averaged
across the 7 series). As we would expect, the variability declines as
the size of the training set increases, though not the average.
\autoref{fig:test-error} shows the average prediction error over the 7
series. It improves markedly as the training set increases to about 400
observations (=100 years) but then plateaus. This is troubling: as we
get more and more data, we can not predict new data any better. This
indicates one of three possibilities: (1) that with about 400
observations, we can estimate the parameters nearly perfectly, (2) that
the model is poorly identified---some parameters will simply never be
well estimated, but we can predict well anyway, or (3) the data are so
highly correlated that the range of training observations we consider is
far too small---we actually need millions of observations in order to
see a meaningful decline in out-of-sample predictive performance. We can
better determine which of these three is occurring by comparing with the
predictive performance of the true parameters and examining the error in
parameter estimates. The blue line in \autoref{fig:test-error} is the
out-of-sample mean prediction error for the true parameters. The test
error is not getting any closer to this ideal scenario, plateauing
slightly above the baseline by about 400 training points. This seems to
suggest that explanation (2) is accurate: even with more data, we will
never be able to recover the true parameters, though we get some
improvement in predictions relatively quickly.

\begin{figure}[t]

{\centering \includegraphics{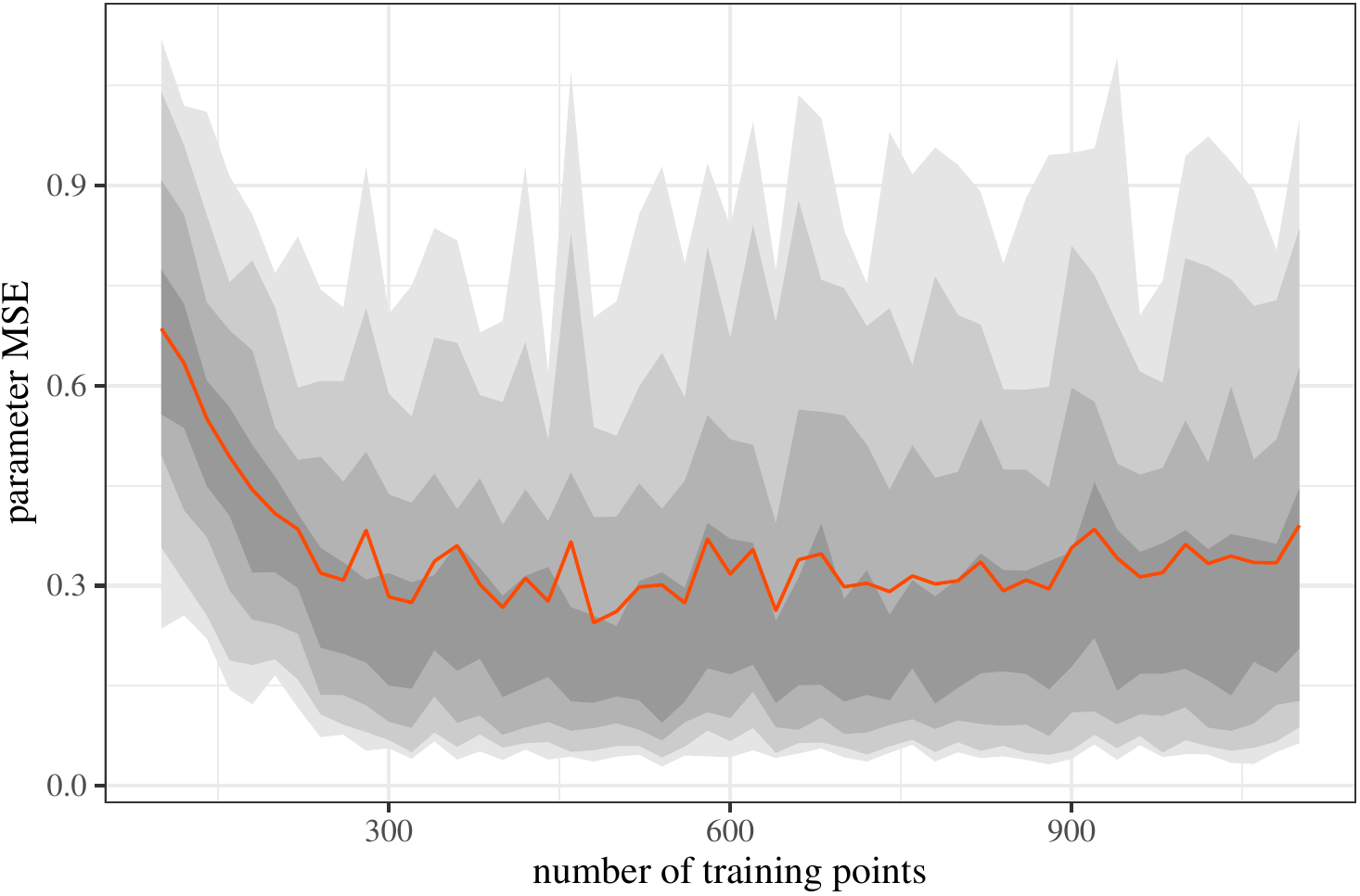} 

}

\caption{The parameter MSE shows similar behavior to the predictive MSE: steep initial decline toward an asymptote greater than that of the true model.}\label{fig:parm-error}
\end{figure}

\begin{figure}[t]

{\centering \includegraphics{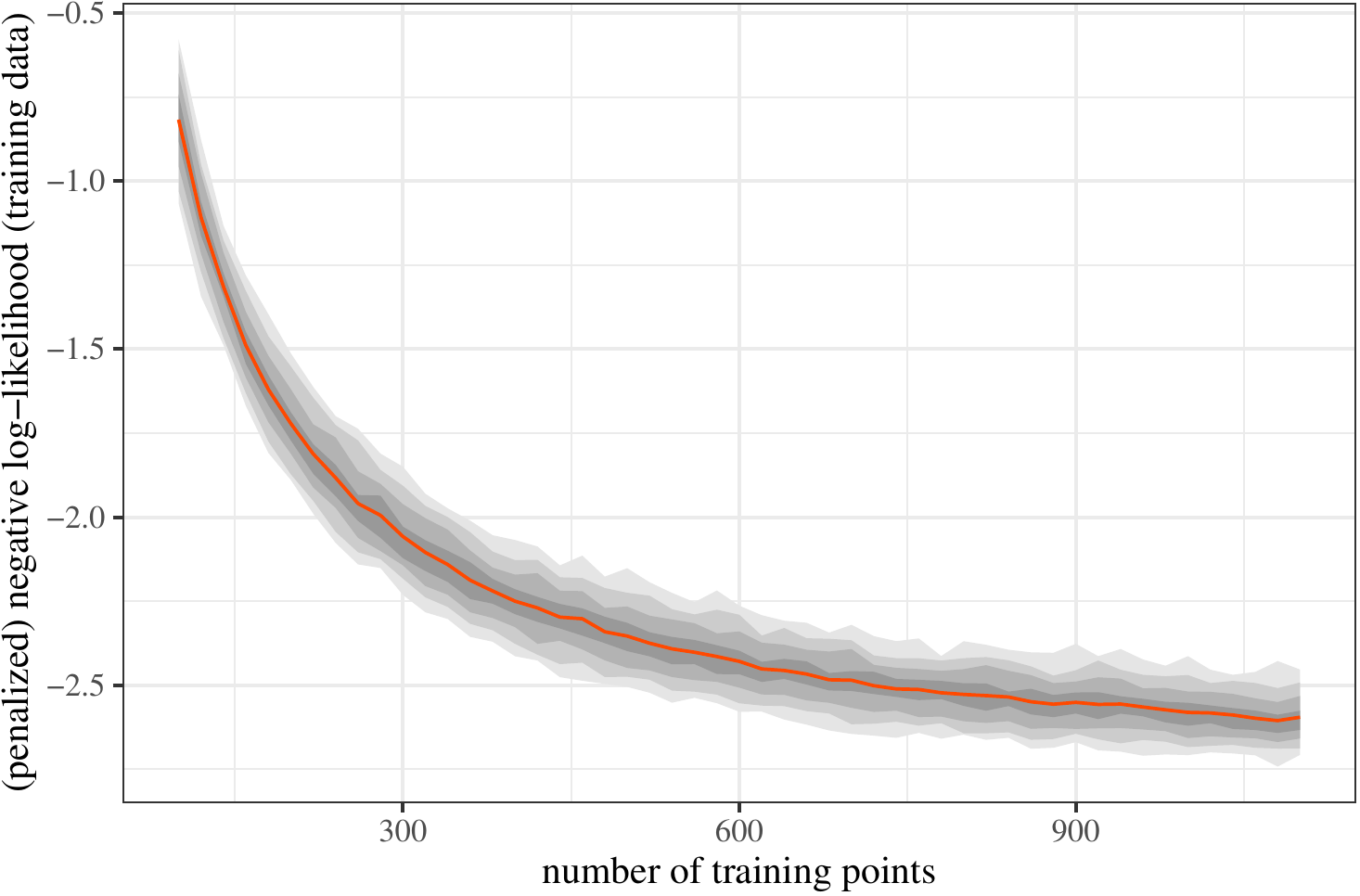} 

}

\caption{The negative log-likelihood (in-sample) per observation.}\label{fig:pen-negll}
\end{figure}

\begin{figure}[t]

{\centering \includegraphics{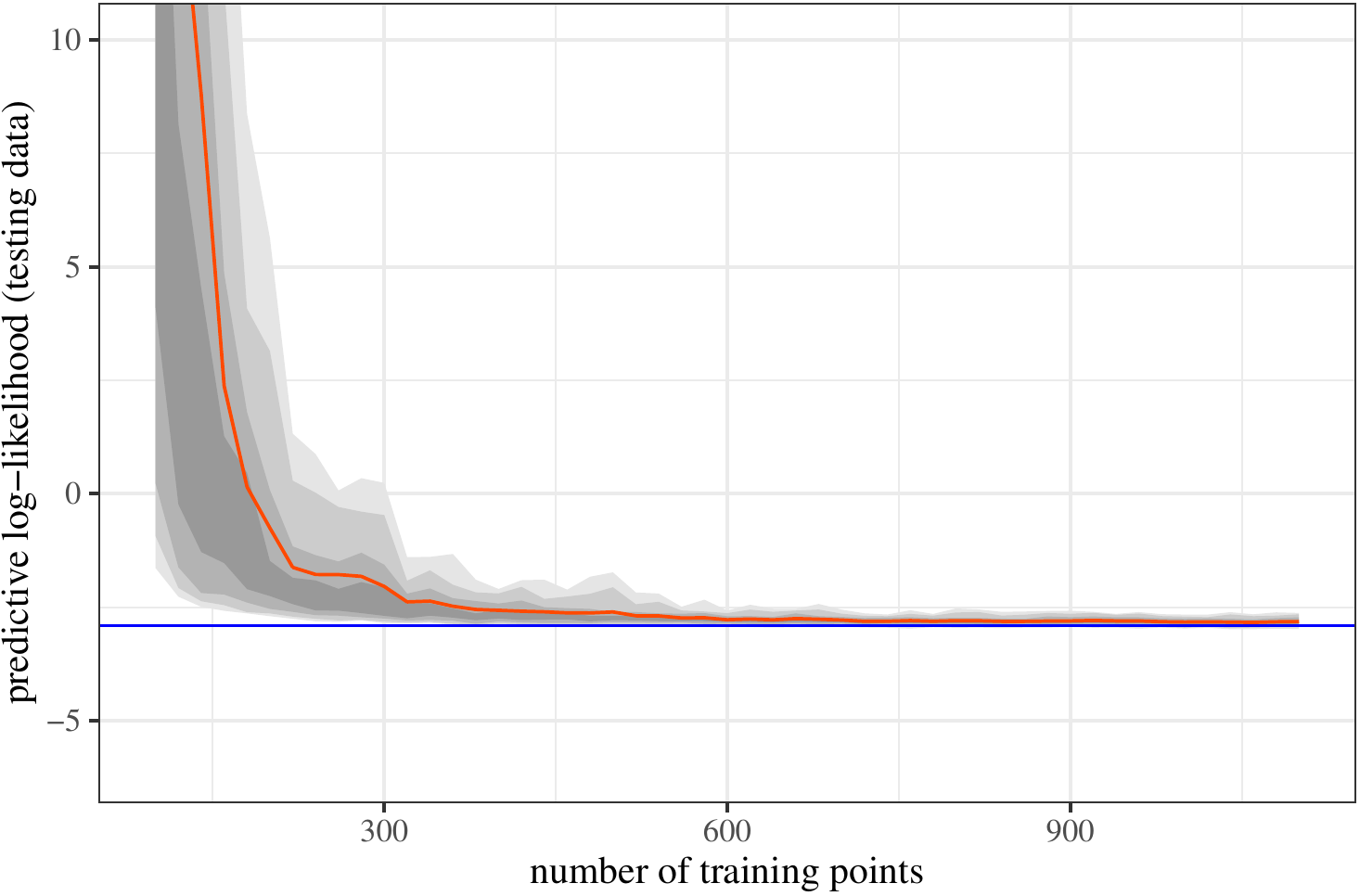} 

}

\caption{The negative predictive log-likelihood (out-of-sample) per observation.}\label{fig:predict-negll}
\end{figure}

\begin{figure}[t]

{\centering \includegraphics{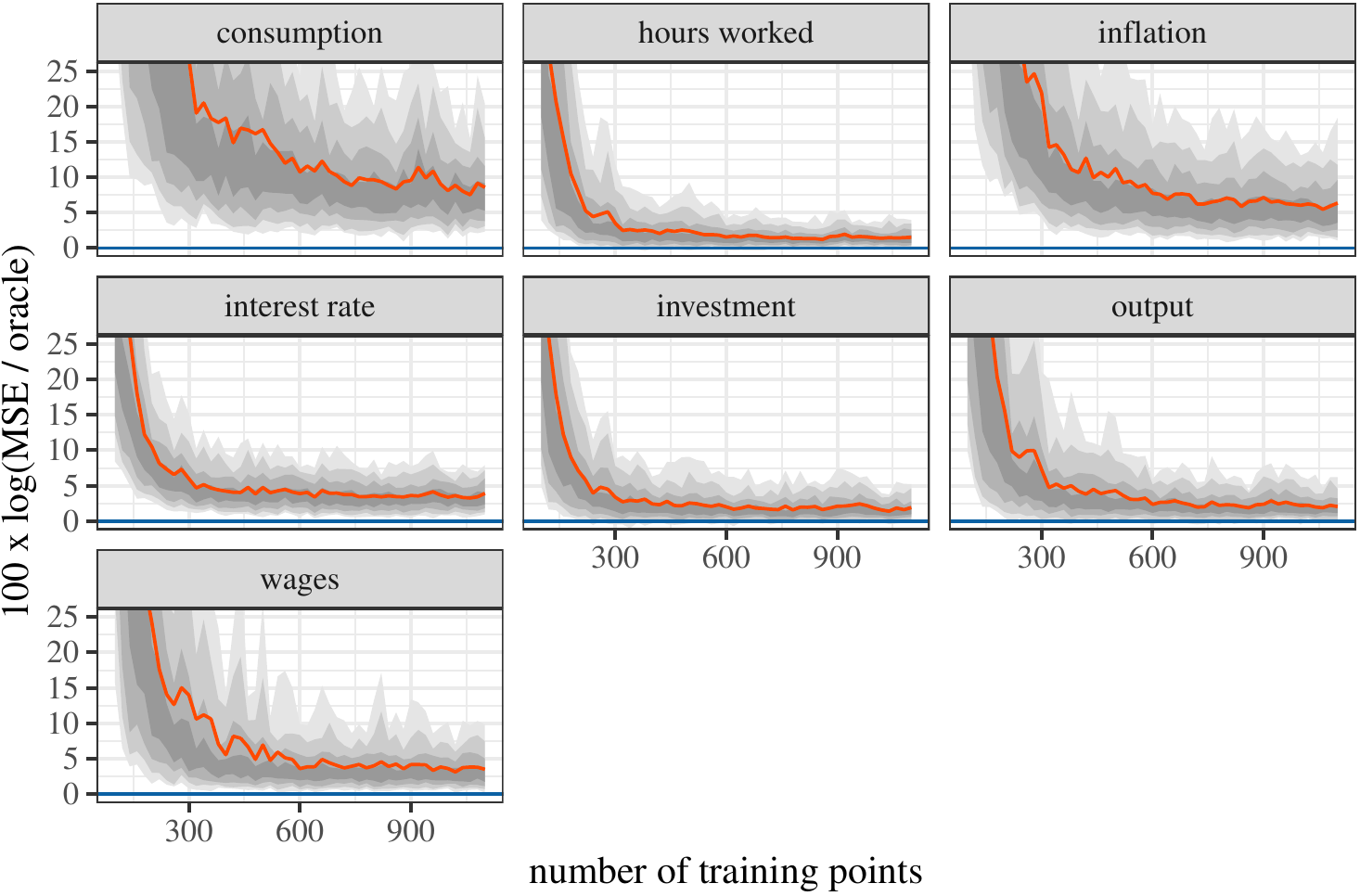} 

}

\caption{Mean-squared prediction error for each series.}\label{fig:individual-series}
\end{figure}

Examining the parameter error serves to confirm this conjecture.
\autoref{fig:parm-error} shows the average squared error of the
parameter estimates. Improvement in this metric also stagnates despite
the growing training sets. \autoref{fig:pen-negll} and
\autoref{fig:predict-negll} show the change in information per
observation (in sample) and per prediction (out of sample). For maximum
likelihood inference, we would expect both of these to decline as the
amount of training data increases. These figures therefore confirm that
the estimation procedure is responding to the increased sample size even
though prediction and estimation errors both fail to improve
meaningfully. \autoref{fig:individual-series} shows the average
prediction error for each series individually. From this decomposition,
prediction of all time series improves, but only labor ever reaches the
minimum (relative to the truth). The other series are all predicted
poorly.

\begin{figure}[t]

{\centering \includegraphics{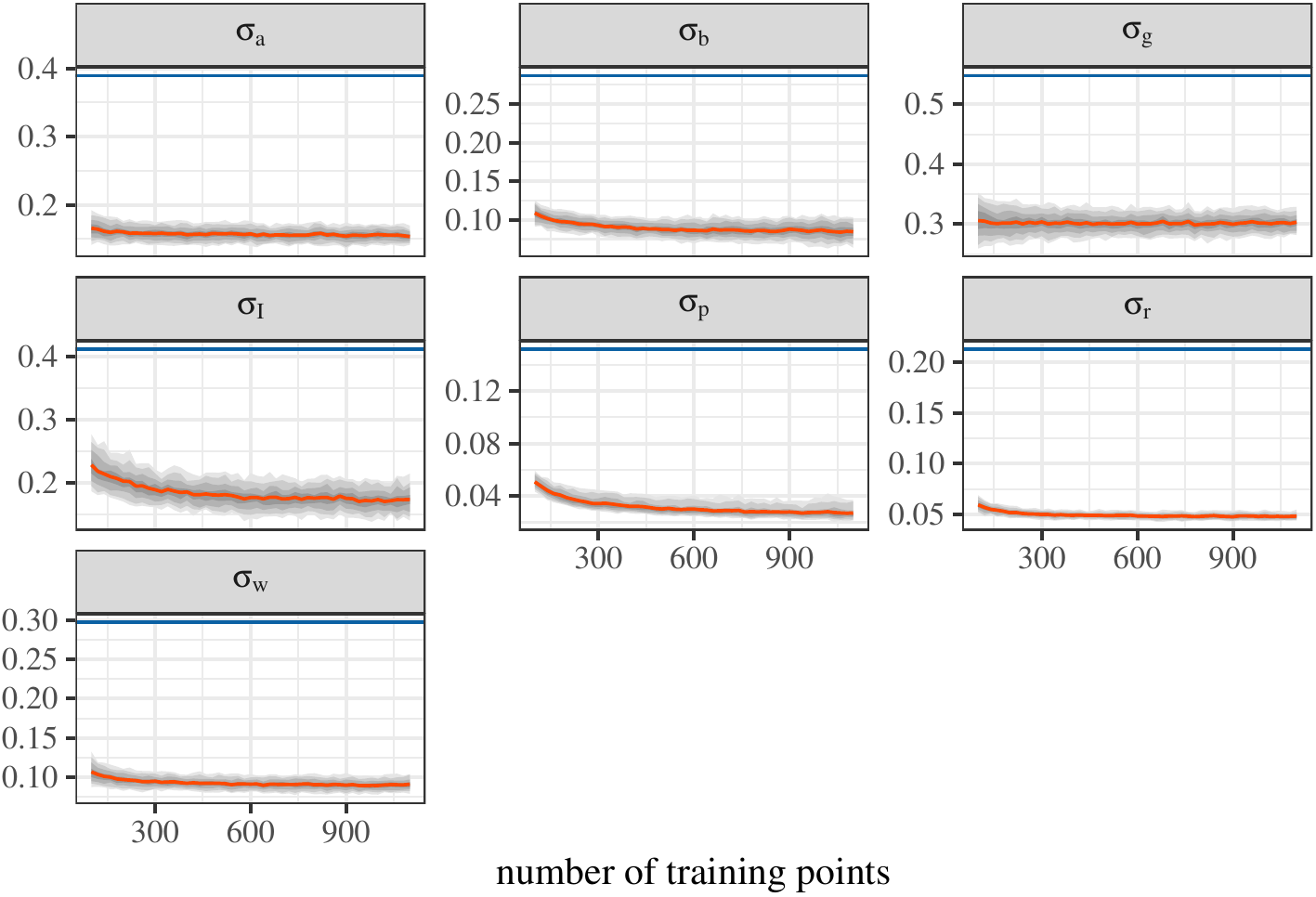} 

}

\caption{Estimates for the standard deviations of stochastic shocks.}\label{fig:shock-sds}
\end{figure}

\begin{figure}[t]

{\centering \includegraphics{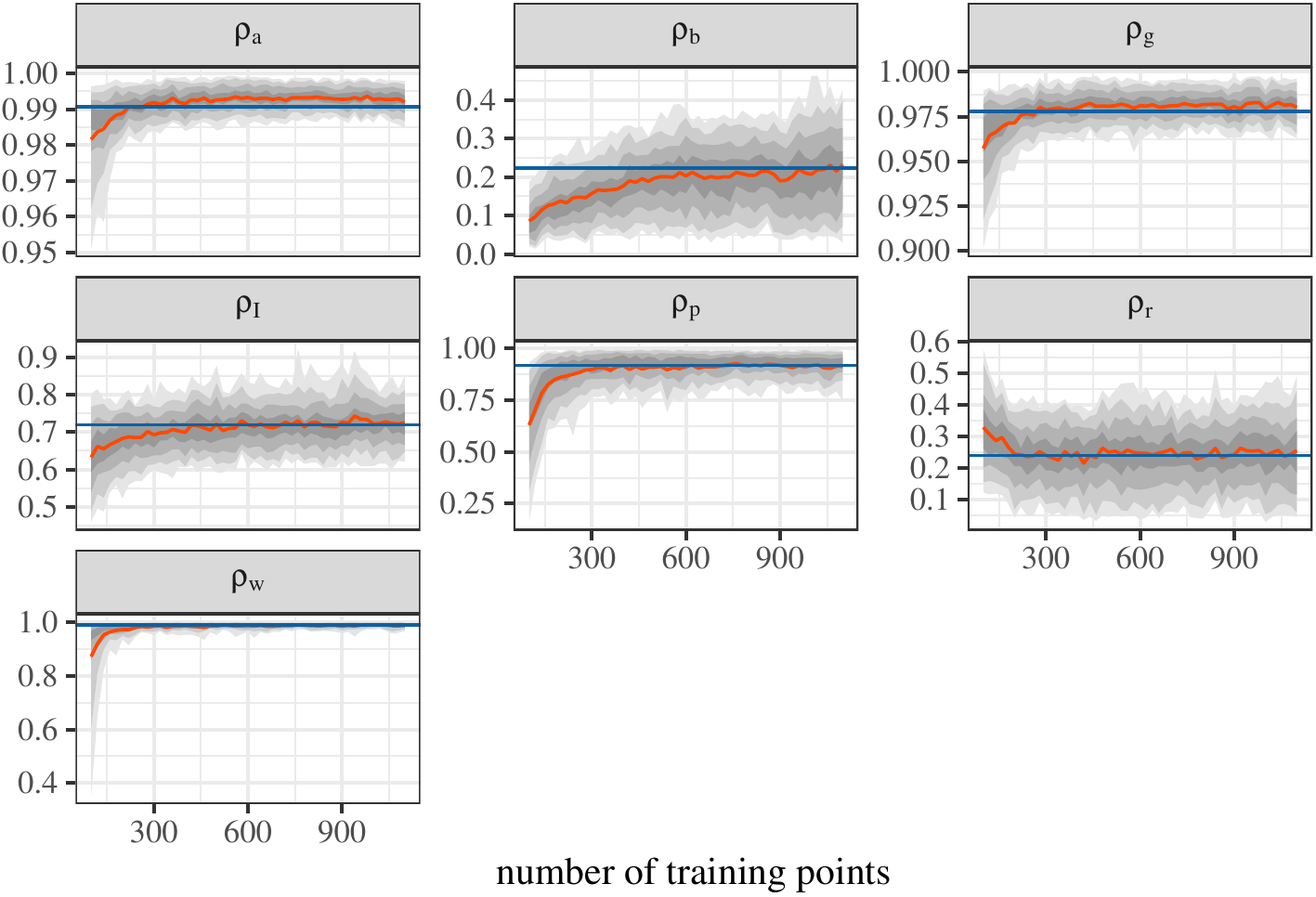} 

}

\caption{Estimates for the autocorrelation parameters in the shock processes.}\label{fig:autocorrelations-shocks}
\end{figure}

\begin{figure}[t]

{\centering \includegraphics{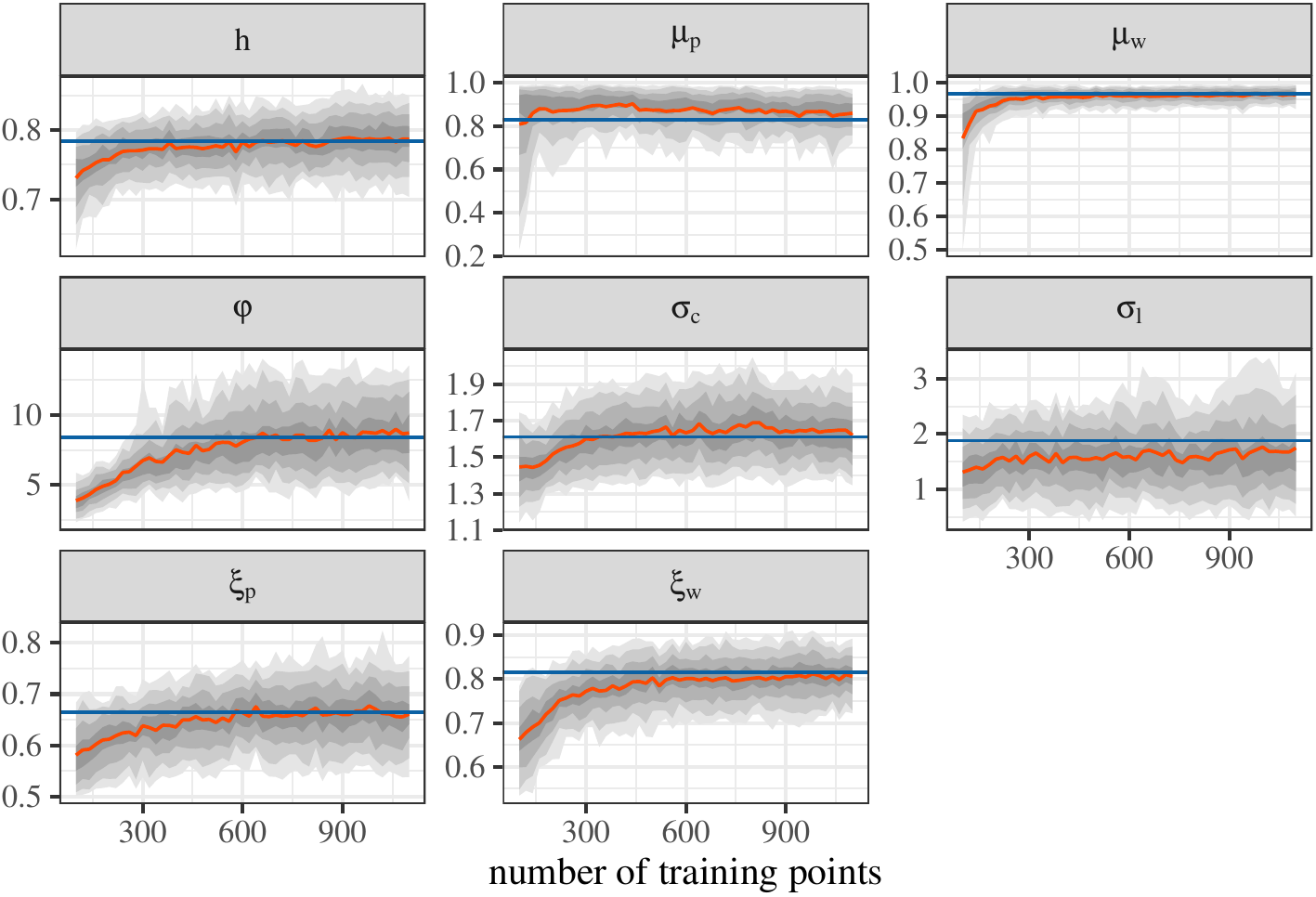} 

}

\caption{Estimates of "deep" parameters}\label{fig:deep1}
\end{figure}

\begin{figure}[t]

{\centering \includegraphics{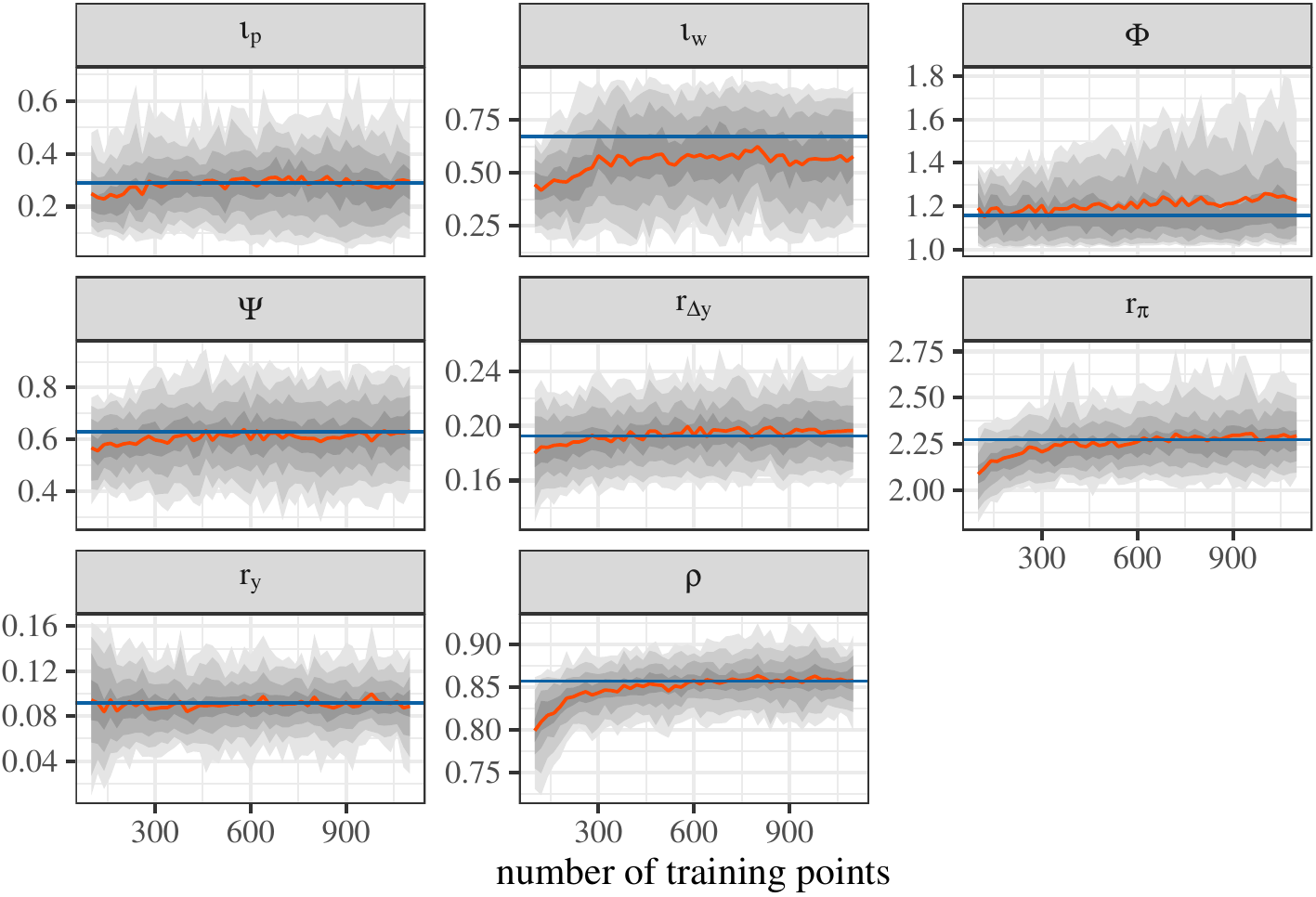} 

}

\caption{Estimates of "deep" parameters}\label{fig:deep2}
\end{figure}

\begin{figure}[t]

{\centering \includegraphics{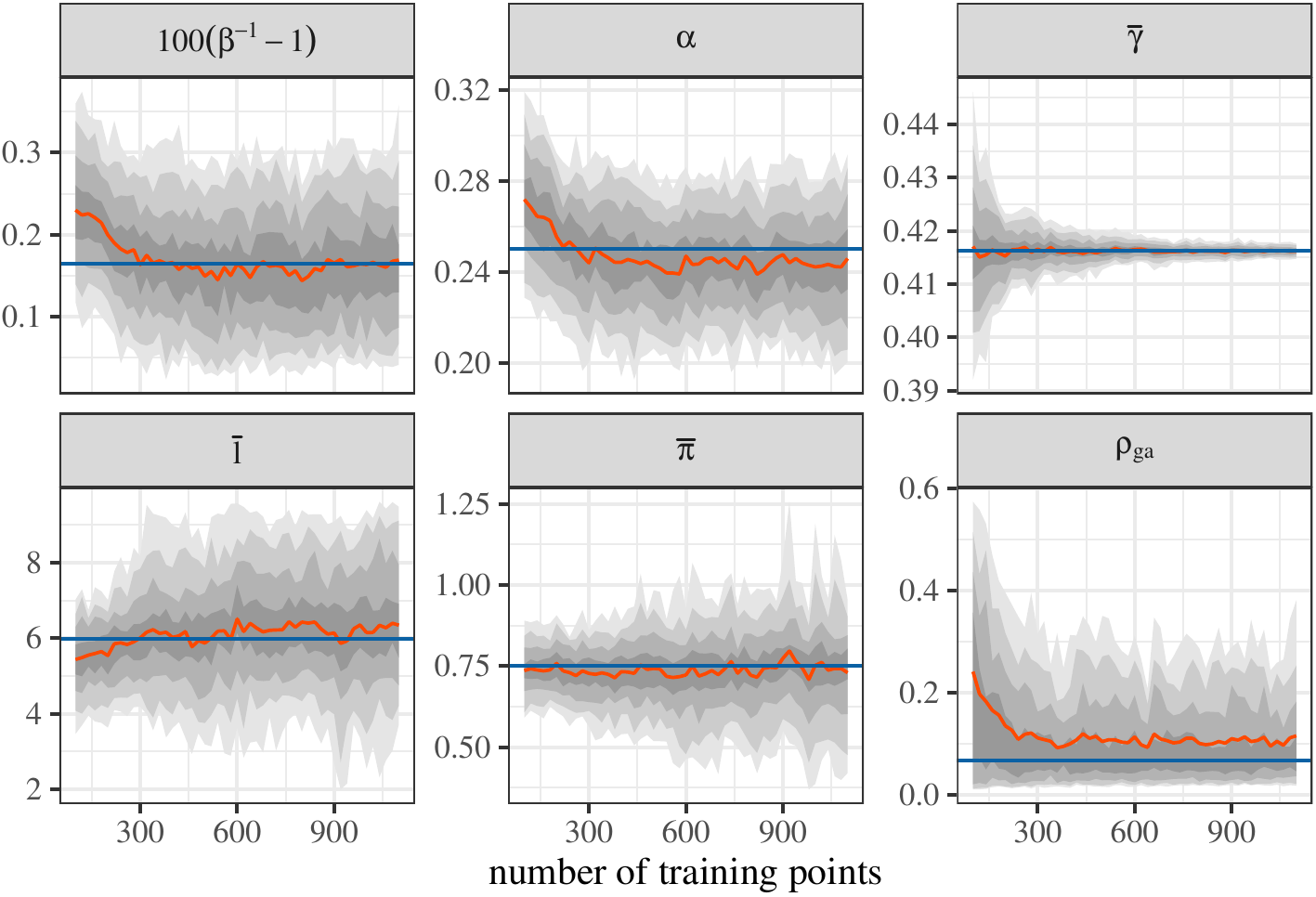} 

}

\caption{Estimates of "deep" parameters}\label{fig:deep3}
\end{figure}

Figures \ref{fig:shock-sds}--\ref{fig:deep3} show the estimates of each
parameter individually. The horizontal red lines indicate the truth.
Here we can see that while many parameters are well-estimated with
decreasing variability as the training sets grow, others converge to the
wrong values, or fail to converge at all. This phenomenon holds
especially for both of the parameters of the shock processes, but also
for some of the economically-relevant ``deep'' parameters. For instance,
\(\sigma_l\), the elasticity of labor supply with respect to the real
wage, is consistently underestimated by about -93\%. The data provides
essentially no information about \(\iota_w\), which measures the
dependence of real wages on lagged inflation. Other parameters which are
poorly estimated include \(\varphi\), the steady-state elasticity of the
capital adjustment cost function. In all these cases, estimation is
biased, so using the estimated values from the real data to draw
conclusions about the real economy is unwise. It is possible that this
bias is due to the linearization procedure\footnote{It is possible,
  though not all that plausible, that a misspecified linearized model
  might be \emph{unbiased} for these parameters inside the original
  nonlinear model.}, in which case, not only should linearized models be
avoided for prediction, but also for drawing any economic inferences.
Overall, most parameters are poorly identified as evidenced by their
stable (rather than decreasing) variability with more data.

To examine case (3)---that the data is so correlated that even 250 years
is too little to produce accurate estimates of the true parameters---we
may ask how closely the information per observation displayed in
\autoref{fig:pen-negll} approaches the theoretical asymptotic
limit.\footnote{We are adapting a procedure recommended by \citet{Andy-Fraser-on-HMMs} for evaluating general hidden Markov models.}
For essentially any stationary ergodic stochastic process, the limit of
the negative loglikelihood per observation as the number of observations
approaches infinity exists and is unique \citep{Gray-entropy-2nd}. This
limit is the \emph{entropy rate} of the process. Mathematically, with
probability 1, \[
\lim_{n\rightarrow\infty} -\frac{1}{n}\ell(X_{1:n}; \theta_0) = h(\theta_0),
\] where \(h(\theta_0)\) is the entropy rate. Furthermore, by the
generalized asymptotic equipartition property
\citep{Algoet-and-Cover-on-AEP}, one can show that for any other
parameter vector \(\theta \neq \theta_0\), with probability 1, \[
\lim_{n\rightarrow\infty} -\frac{1}{n}\ell(X_{1:n}; \theta) = h(\theta_0) + \mathrm{KL}(\theta_0\ ||\ \theta),
\] where the second term is the Kullback-Leibler divergence rate
\citep{Gray-entropy-2nd}. Because
\(\mathrm{KL}(\theta_0\ ||\ \theta) \geq 0\), asymptotically,
\(h(\theta_0)\) is a lower bound for the average negative loglikelihood
of any parameter vector.

\begin{figure}[t]

{\centering \includegraphics{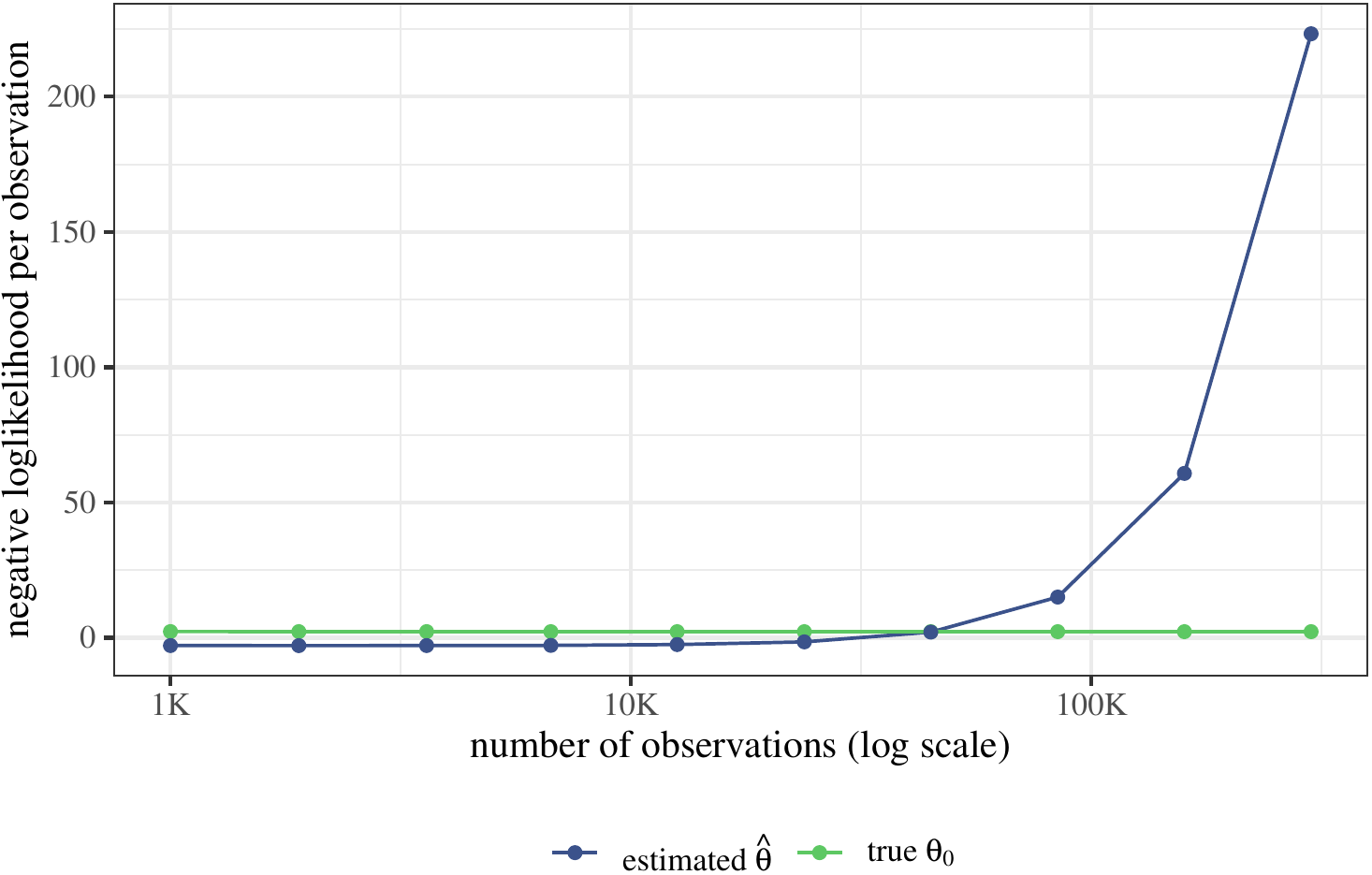} 

}

\caption{Comparing the in-sample negative loglikelihood per observation of a parameter estimated with 1000 observations relative to the true parameter. In the long-run, the entropy rate is a lower bound, but only after about 12,000 years of quarterly data.}\label{fig:entropy-investigation}
\end{figure}

Because the data is generated from a linear Gaussian state space model,
it is not hard to calculate that \(h(\theta_0) \approx 2.25\),
significantly \emph{larger} than any of the values in
\autoref{fig:pen-negll}. Somehow, the estimated parameters generally
have \emph{lower} negative loglikelihoods than the true parameter. The
fact that the standard deviations of the shocks
(\autoref{fig:shock-sds}) are consistently underestimated suggests that,
even over this period, the in-sample fit to the data gives an overly
optimistic picture of the long run behavior: we have not allowed \(n\)
to get large enough. \autoref{fig:entropy-investigation} shows the
negative loglikelihood per observation over a much longer horizon---up
to 75,000 years of quarterly data. For thousands of years, the parameter
vector estimated over the first 250 years is less than the theoretical
lower bound, but eventually, it starts to curve upward. This suggests
that, both explanations (2) and (3) are accurate: in order to estimate a
DSGE accurately, we need massive amounts of data.

\clearpage

\hypertarget{sec:permutation-summary}{%
\section{Permuting the data}\label{sec:permutation-summary}}

A second way to assess the predictive ability (and economic content) of
the SW DSGE model is to perform a simple permutation test, permuting
\emph{across} the series rather than within them. That is, rather than
giving the model data in the order it expects, we swap the data series
with each other and see if the model predicts future data any better. We
estimated the model on the properly ordered data (presented in the
previous section) as well as all 5039 other permutations of the 7 data
series. For each estimation, we used the same estimation procedure as
before to minimize the (penalized) negative log likelihood. The model is
trained using the first 200 time points and its predictive performance
is tested on the remaining 51.

\begin{figure}[t]

{\centering \includegraphics{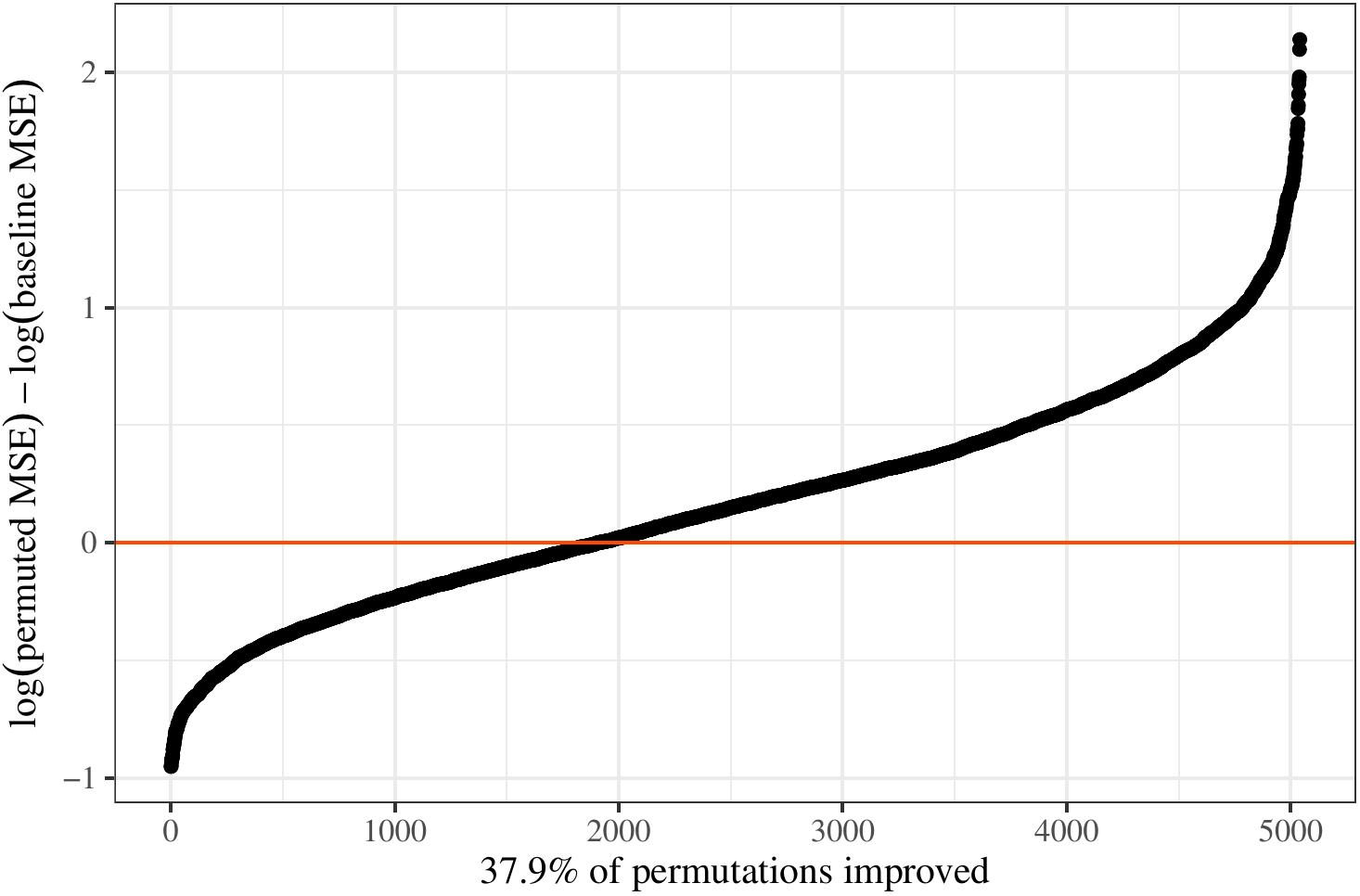} 

}

\caption{Average percentage improvement in out-of-sample forecast MSE. The horizontal red line represents baseline performance. Lower values are better.}\label{fig:mean-percent-improvement}
\end{figure}

\begin{figure}[t]

{\centering \includegraphics{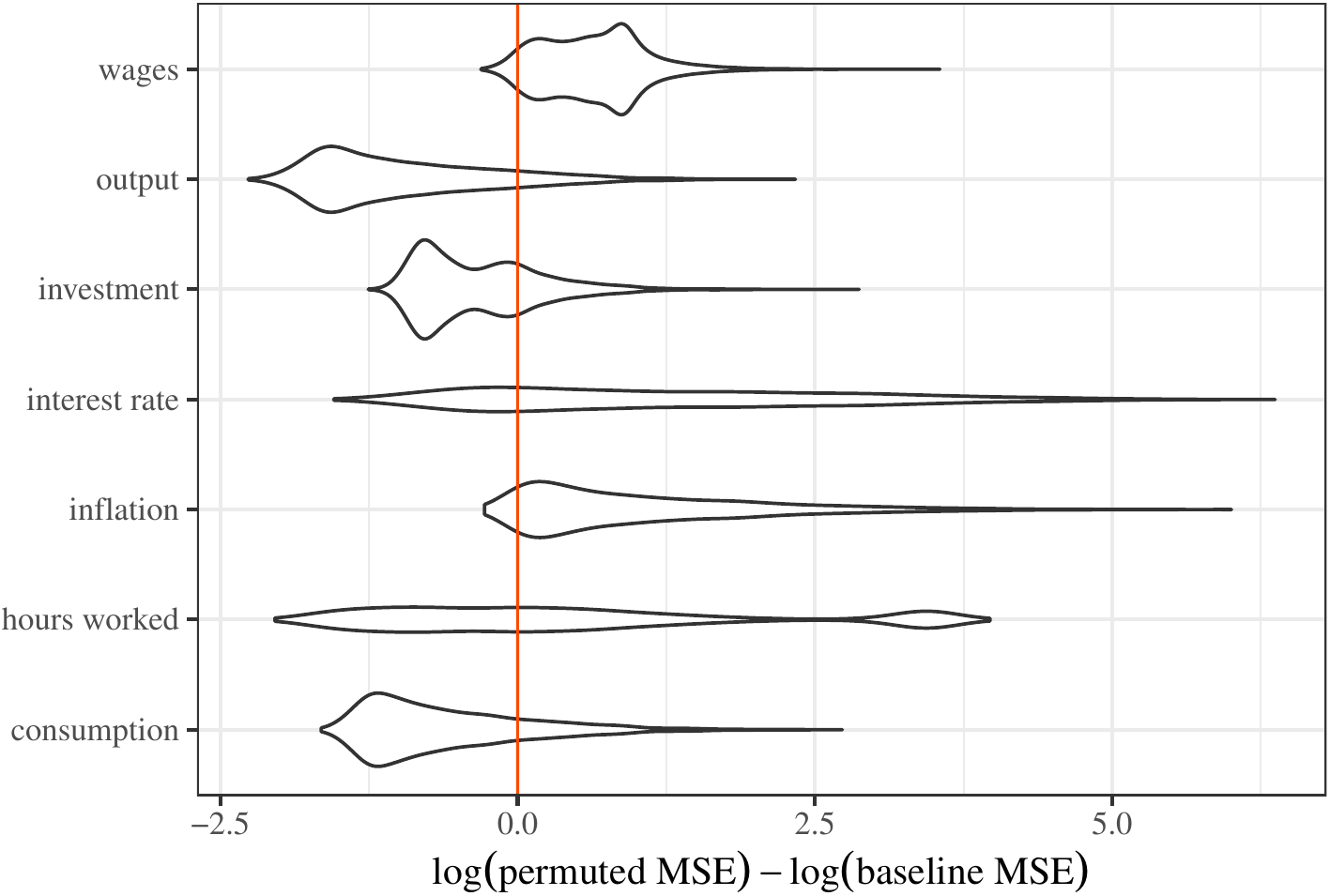} 

}

\caption{Percentage improvement in out-of-sample forecast MSE for each data series individually. The horizontal red line represents baseline performance. Values to the left are better.}\label{fig:series-percent-improvement}
\end{figure}

The next few figures summarize the results. We report three criteria for
measuring relative performance. The first is the percent improvement in
(out-of-sample) mean-squared test error (MSE) calculated as the natural
logarithm of the test error for a particular permutation divided by that
of the baseline model and then averaged across all 7 series:
\begin{equation}
  \mbox{Average Percent MSE Improvement (p)} = \frac{1}{7} \sum_{i=1}^7 \log
  \left(\frac{\sum_{t=201}^{251} (\hat{x}_{it}^{(p)}-x_{it})^2}
    {\sum_{t=201}^{251} (\hat{x}_{it}^{b}-x_{it})^2}\right),
\end{equation} where a superscript \(b\) represents the baseline model
and \((p)\) the \(p^{th}\) permutation. The result is shown in
\autoref{fig:mean-percent-improvement}.
\autoref{fig:series-percent-improvement} shows boxplots for the
percentage improvement separately for each time series. The best model
had an average percentage improvement of 95\% relative to the baseline
model. The vertical line at zero represents performance equivalent to
baseline. About 38\% of permutations had better predictive performance
than the baseline model.

\begin{figure}[t]

{\centering \includegraphics{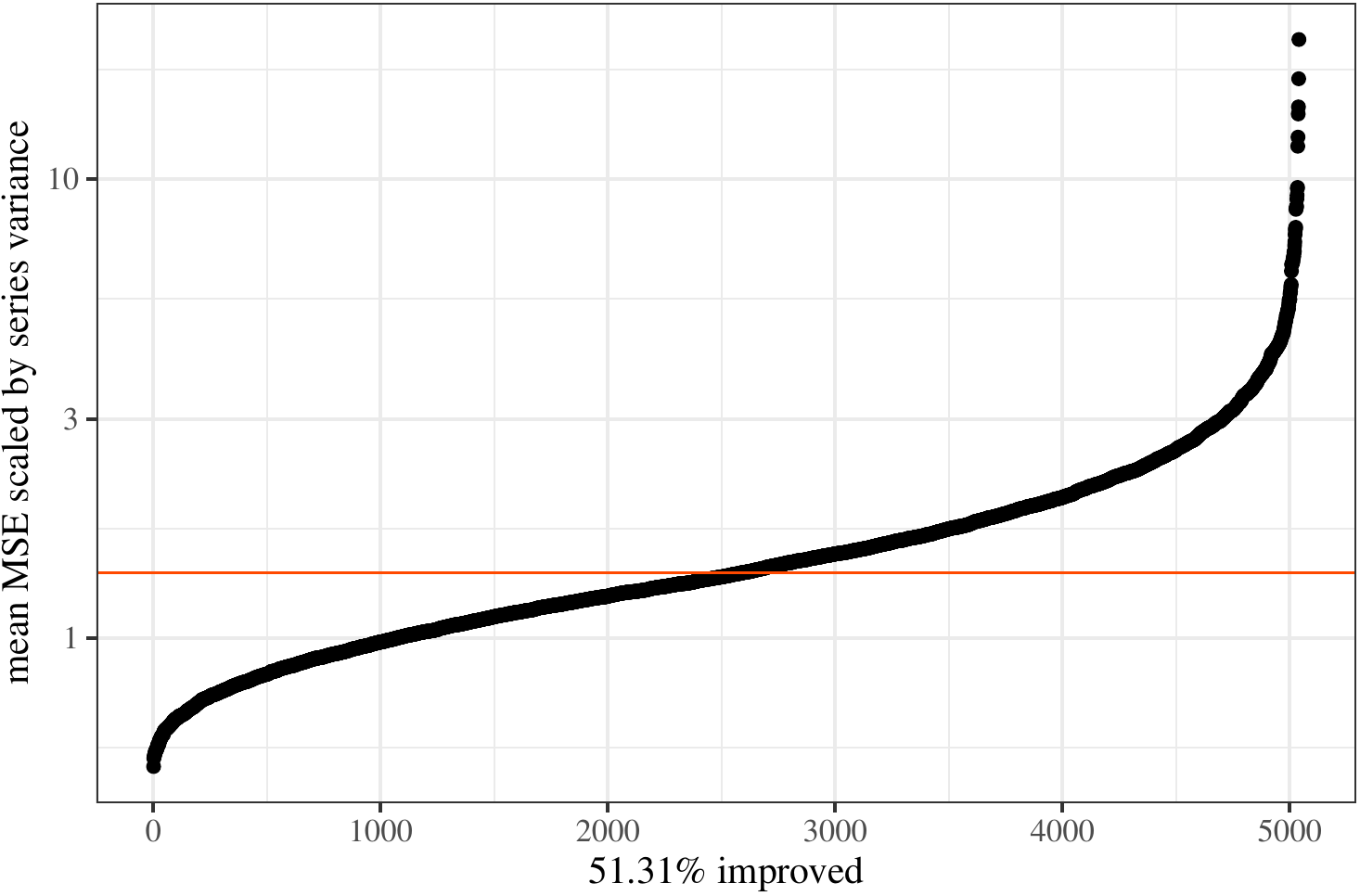} 

}

\caption{Average improvement in out-of-sample forecast MSE scaled by the variance of the data. The horizontal red line represents baseline performance.}\label{fig:mean-scaled-mse}
\end{figure}

\begin{figure}[t]

{\centering \includegraphics{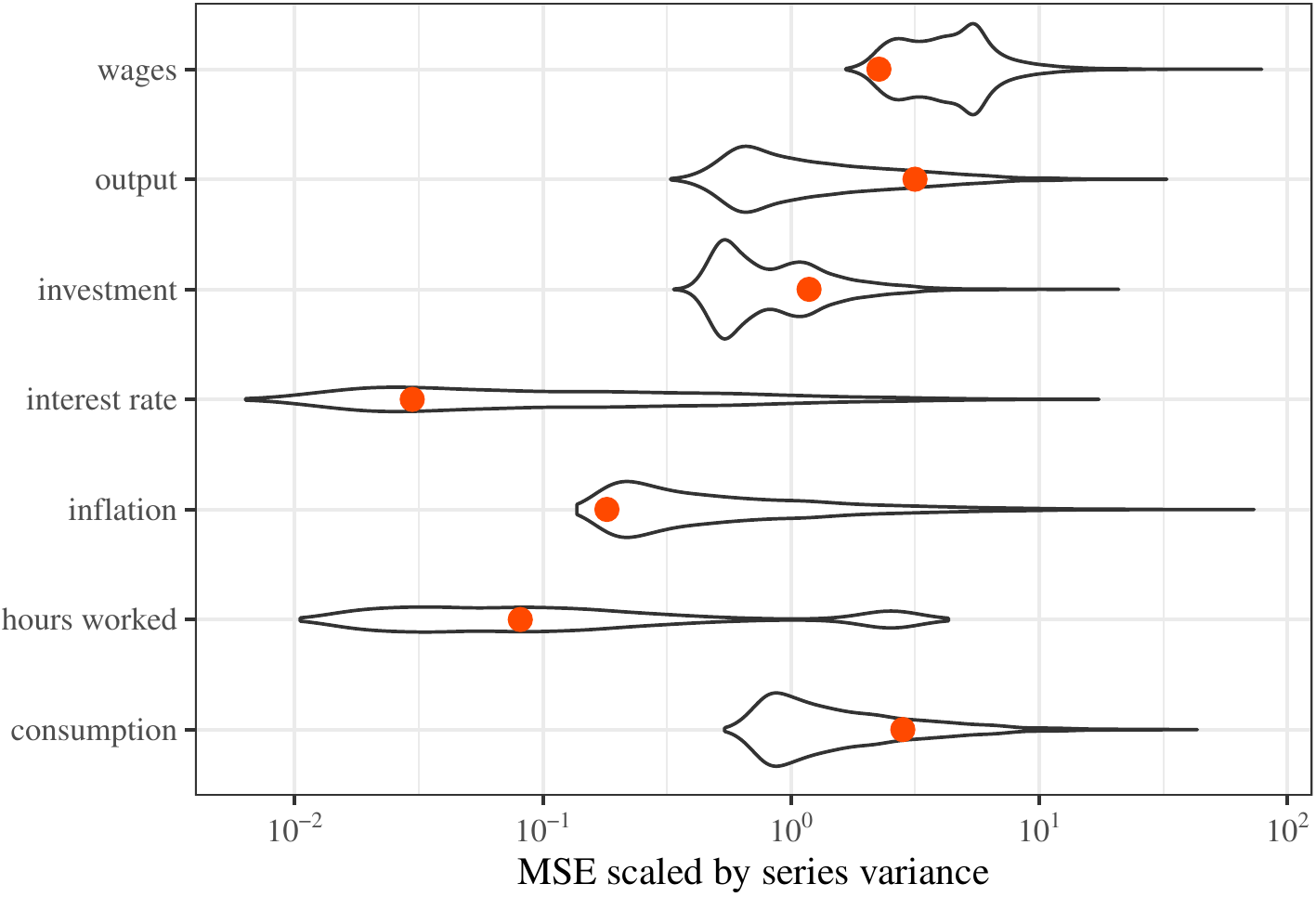} 

}

\caption{Improvement in out-of-sample forecast MSE scaled by the variance of the data for each data series individually. The red dots represents baseline performance.}\label{fig:series-scaled-mse}
\end{figure}

The second measure of performance is simply the out-of-sample MSE scaled
by the observed variance and then averaged across the 7 series:
\begin{equation}
  \mbox{Average Scaled MSE (p)} = \frac{1}{7} \sum_{i=1}^7
  \frac{\sum_{t=201}^{251} (\hat{x}_{it}^{(p)}-x_{it})^2}
    {\mbox{Var}(x_i)}.
\end{equation} \autoref{fig:mean-scaled-mse} displays the average of
this measure across all 251 series. Again, about 51\% of permutations
achieved better average scaled MSE than the baseline model. The best
permutation achieved an average scaled MSE of 0.53 relative to 1.39 for
the baseline model. \autoref{fig:series-scaled-mse} shows boxplots for
the scaled MSE separately for each series. Red dots indicate the
performance of the baseline model.

\begin{figure}[t]

{\centering \includegraphics{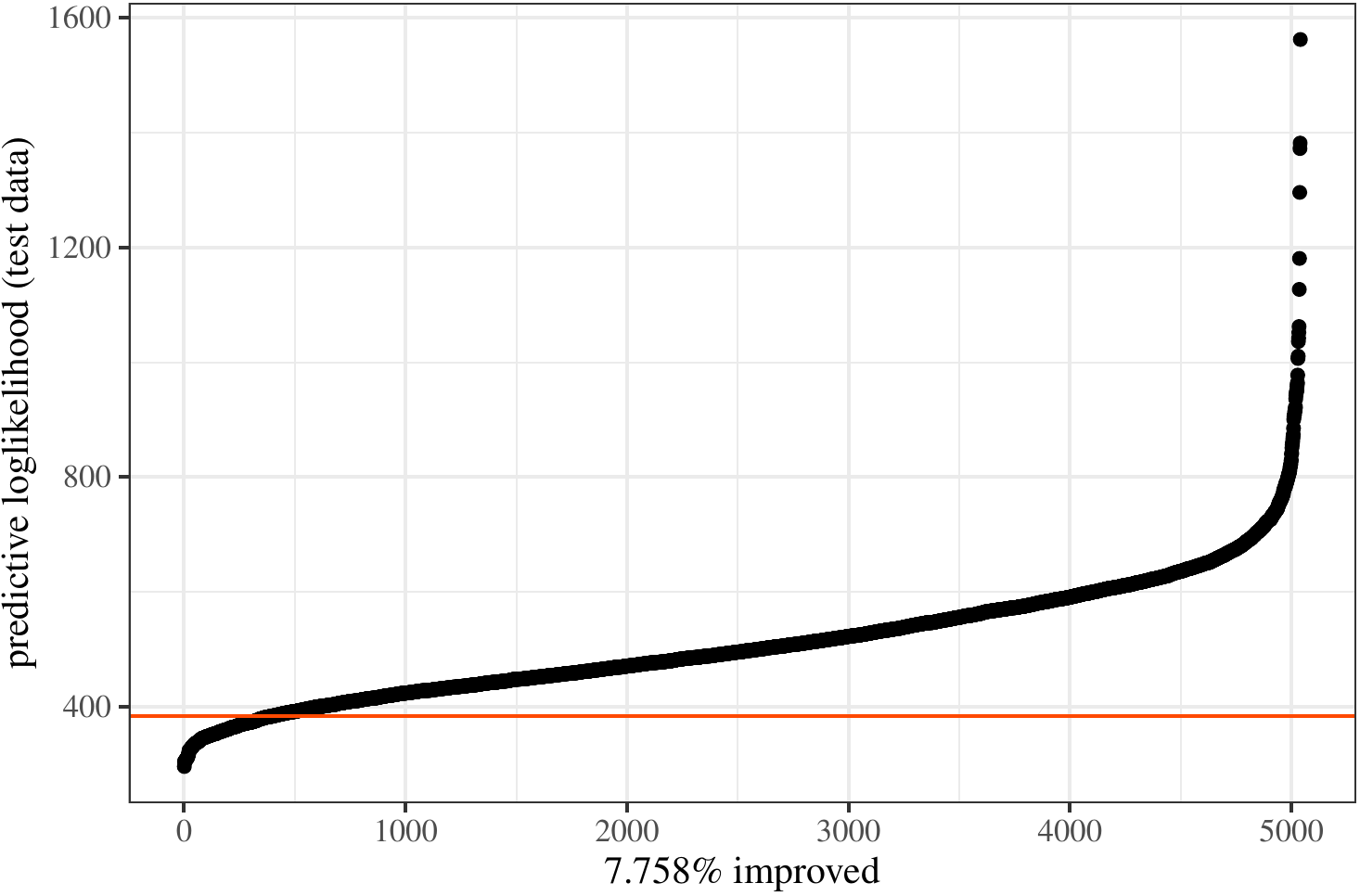} 

}

\caption{Negative log predictive likelihood for each model. The horizontal red line represents baseline performance. Note that this is an out-of-sample performance measure.}\label{fig:loglike-pred-ml}
\end{figure}

A total of 0 of permutations resulted in a lower negative penalized log
likelihood. A Bayesian interpretation of these results would be that 0
of the permuted ``models'' are preferable to the true, unpermuted,
economic model. It should be noted that this is an in-sample measure of
performance and that the Bayesian interpretation is conditional on the
model being true and the priors accurately reflecting expert
information. An out-of-sample evaluation (without penalty) shows that
8\% of the premuted models actually have better prediction performance
when evaluated through the likelihood. \autoref{fig:loglike-pred-ml}
gives a visual depiction.

\begin{table}

\caption{\label{tab:best-results}Series MSEs for SW and top models.}
\centering
\resizebox{\linewidth}{!}{
\begin{tabular}[t]{rrrrrrr}
\toprule
hours worked & interest rate & inflation & output & consumption & investment & wages\\
\midrule
1.12 & 0.02 & 0.06 & 2.32 & 1.36 & 5.40 & 1.06\\
0.24 & 0.01 & 0.06 & 0.39 & 0.34 & 2.64 & 0.84\\
0.25 & 0.02 & 0.09 & 0.33 & 0.41 & 1.56 & 0.97\\
0.32 & 0.01 & 0.09 & 0.42 & 0.35 & 2.07 & 1.33\\
\bottomrule
\end{tabular}}
\end{table}

\begin{table}

\caption{\label{tab:top-20-perc}Permutations with highest \% improvement}
\centering
\resizebox{\linewidth}{!}{
\begin{tabular}[t]{llllllll}
\toprule
hours worked & interest rate & inflation & output & consumption & investment & wages & \# different\\
\midrule
investment & hours worked & interest rate & wages & output & inflation & consumption & 7\\
hours worked & interest rate & consumption & wages & inflation & output & investment & 5\\
hours worked & interest rate & output & wages & inflation & consumption & investment & 5\\
hours worked & investment & inflation & wages & interest rate & consumption & output & 5\\
hours worked & interest rate & wages & output & inflation & consumption & investment & 4\\
\addlinespace
consumption & hours worked & interest rate & wages & output & investment & inflation & 6\\
hours worked & investment & interest rate & wages & inflation & output & consumption & 6\\
inflation & hours worked & interest rate & output & consumption & investment & wages & 3\\
consumption & hours worked & inflation & wages & interest rate & output & investment & 6\\
inflation & hours worked & interest rate & consumption & output & investment & wages & 5\\
\addlinespace
hours worked & interest rate & wages & consumption & inflation & output & investment & 5\\
investment & hours worked & inflation & output & interest rate & consumption & wages & 4\\
hours worked & inflation & consumption & output & wages & investment & interest rate & 4\\
consumption & hours worked & interest rate & wages & inflation & output & investment & 7\\
hours worked & inflation & interest rate & investment & output & consumption & wages & 5\\
\addlinespace
hours worked & inflation & output & consumption & wages & investment & interest rate & 5\\
inflation & hours worked & interest rate & output & investment & wages & consumption & 6\\
hours worked & inflation & consumption & wages & interest rate & output & investment & 6\\
investment & hours worked & interest rate & wages & consumption & inflation & output & 6\\
inflation & hours worked & interest rate & investment & output & wages & consumption & 7\\
\bottomrule
\end{tabular}}
\end{table}

\begin{table}

\caption{\label{tab:top-20-scmse}Permutations with lowest average scaled MSE}
\centering
\resizebox{\linewidth}{!}{
\begin{tabular}[t]{llllllll}
\toprule
hours worked & interest rate & inflation & output & consumption & investment & wages & \# different\\
\midrule
hours worked & inflation & consumption & wages & interest rate & output & investment & 6\\
hours worked & interest rate & output & wages & inflation & consumption & investment & 5\\
hours worked & interest rate & consumption & wages & inflation & output & investment & 5\\
hours worked & investment & consumption & wages & interest rate & output & inflation & 6\\
hours worked & interest rate & wages & output & inflation & consumption & investment & 4\\
\addlinespace
consumption & hours worked & inflation & wages & interest rate & output & investment & 6\\
hours worked & interest rate & wages & consumption & inflation & output & investment & 5\\
hours worked & inflation & consumption & output & interest rate & investment & wages & 3\\
hours worked & output & inflation & wages & interest rate & consumption & investment & 5\\
inflation & hours worked & consumption & wages & output & interest rate & investment & 7\\
\addlinespace
hours worked & investment & consumption & wages & output & interest rate & inflation & 6\\
hours worked & investment & output & wages & consumption & interest rate & inflation & 5\\
investment & hours worked & inflation & output & interest rate & consumption & wages & 4\\
hours worked & inflation & interest rate & wages & consumption & output & investment & 5\\
consumption & hours worked & interest rate & wages & inflation & output & investment & 7\\
\addlinespace
hours worked & consumption & interest rate & wages & inflation & investment & output & 5\\
consumption & hours worked & inflation & wages & output & interest rate & investment & 6\\
hours worked & consumption & inflation & wages & interest rate & output & investment & 5\\
hours worked & output & interest rate & consumption & inflation & investment & wages & 4\\
hours worked & investment & output & wages & inflation & interest rate & consumption & 6\\
\bottomrule
\end{tabular}}
\end{table}

\begin{table}

\caption{\label{tab:top-20-llike}Permutations with lowest negative log likelihood}
\centering
\resizebox{\linewidth}{!}{
\begin{tabular}[t]{llllllll}
\toprule
hours worked & interest rate & inflation & output & consumption & investment & wages & \# different\\
\midrule
hours worked & interest rate & output & wages & inflation & investment & consumption & 4\\
hours worked & interest rate & consumption & wages & inflation & investment & output & 4\\
hours worked & interest rate & consumption & wages & inflation & output & investment & 5\\
investment & hours worked & interest rate & wages & output & inflation & consumption & 7\\
interest rate & hours worked & inflation & wages & consumption & investment & output & 4\\
\addlinespace
inflation & hours worked & interest rate & consumption & output & investment & wages & 5\\
inflation & hours worked & interest rate & wages & output & investment & consumption & 6\\
inflation & hours worked & interest rate & wages & consumption & investment & output & 5\\
inflation & hours worked & interest rate & output & consumption & investment & wages & 3\\
hours worked & inflation & interest rate & investment & output & consumption & wages & 5\\
\addlinespace
hours worked & consumption & interest rate & wages & inflation & investment & output & 5\\
hours worked & interest rate & output & wages & inflation & consumption & investment & 5\\
hours worked & inflation & interest rate & wages & consumption & investment & output & 4\\
consumption & hours worked & interest rate & wages & output & investment & inflation & 6\\
hours worked & output & interest rate & wages & inflation & investment & consumption & 5\\
\addlinespace
interest rate & hours worked & inflation & wages & output & investment & consumption & 5\\
hours worked & consumption & output & wages & inflation & investment & interest rate & 5\\
interest rate & hours worked & inflation & wages & output & consumption & investment & 6\\
inflation & hours worked & interest rate & wages & investment & consumption & output & 7\\
hours worked & inflation & consumption & wages & interest rate & output & investment & 6\\
\bottomrule
\end{tabular}}
\end{table}

\autoref{tab:best-results} shows the error measures separately for each
series of best permutations (as measured by out-of-sample performance)
relative to that of the model fit to the true data. Note that the best
permutation is different for the two measures. Tables
\ref{tab:top-20-perc}--\ref{tab:top-20-llike} show the 20 best
permutations for each evaluation method. There are few commonalities
across the best models. One aspect to note is that output shows up in
many places \emph{except} where it is supposed to go. Many models
perform better with wages in that position. It is unclear whether this
is because the model is exceptionally bad at predicting output, because
that slot likes to predict series with small variance, or because that
slot is simply over-regularized. Note that hours worked, and to lesser
extent, the interest rate, are the only series that appear consistently
in the correct places.

\begin{figure}[t]

{\centering \includegraphics{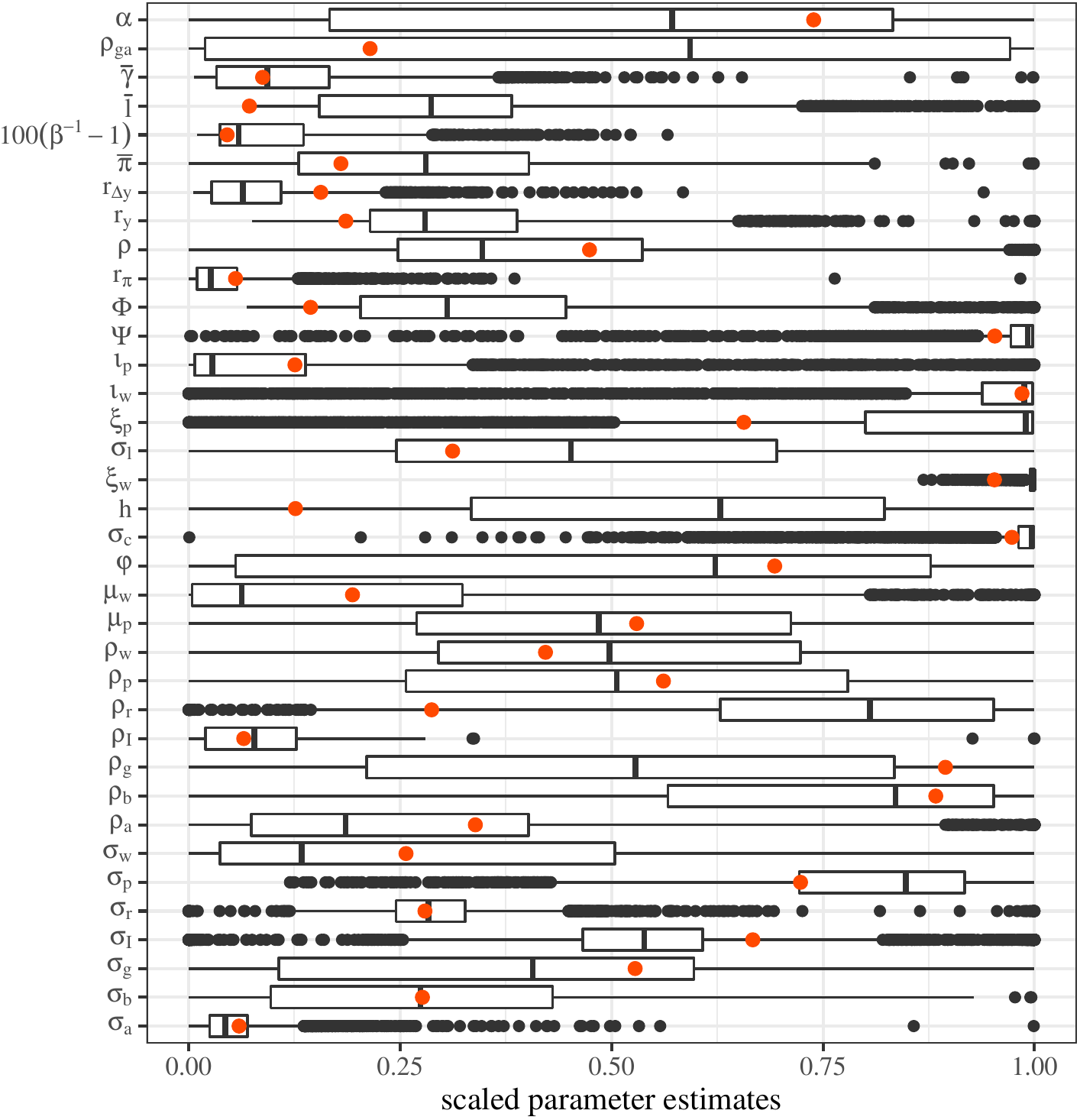} 

}

\caption{Parameter estimates rescaled to [0,1] based on the prior limits. Red points indicate the SW estimates. Black points are outliers relative to the bulk of the permuted estimates.}\label{fig:scaled-parameter-boxplots}
\end{figure}

\begin{figure}[t]

{\centering \includegraphics{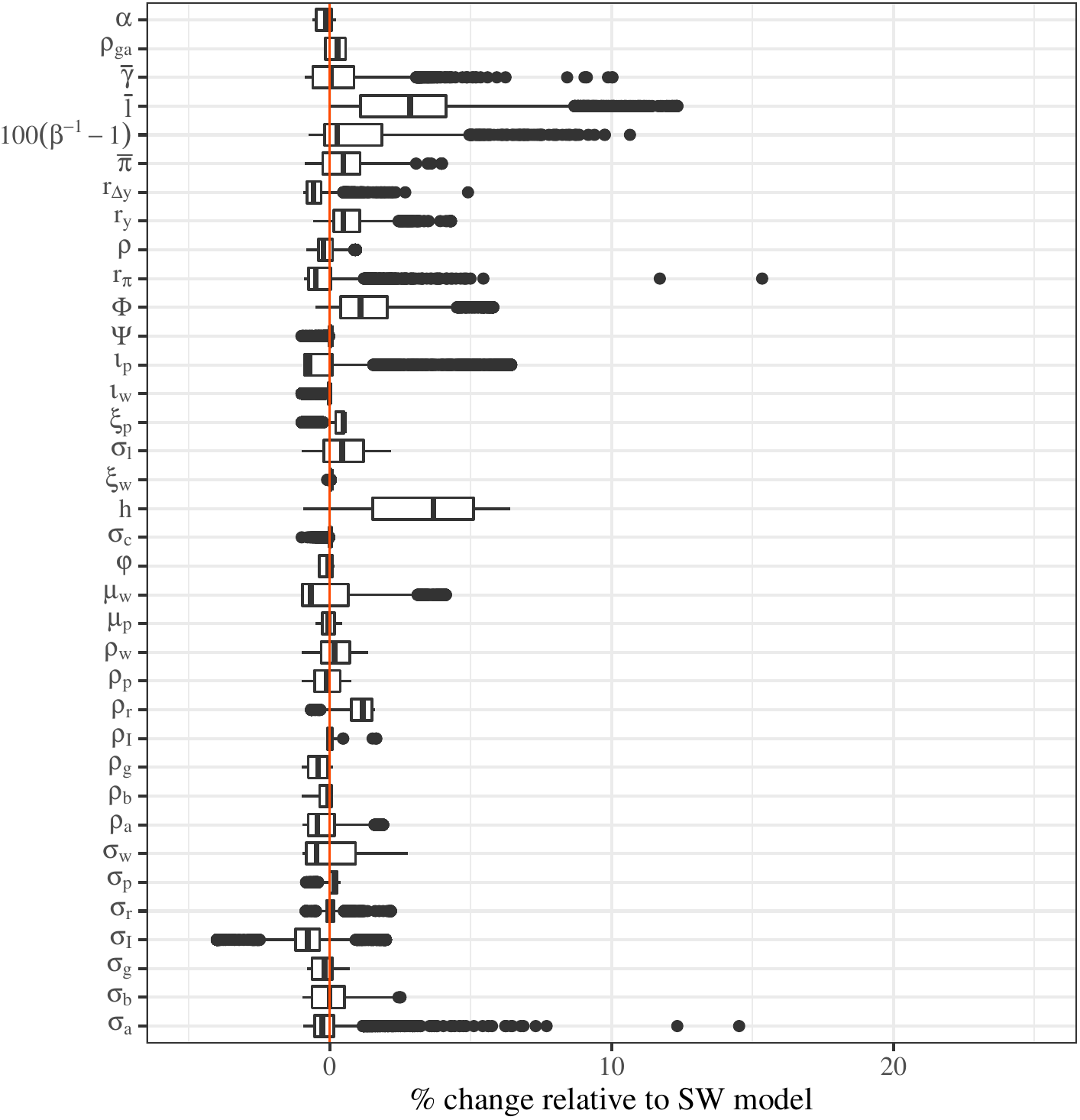} 

}

\caption{Percentage change in estimated parameter relative to the SW estimates. Some large outliers have been removed to better show the bulk.}\label{fig:perc-parameter-deviation}
\end{figure}

\autoref{fig:scaled-parameter-boxplots} shows boxplots for the parameter
estimates across permutations scaled to \([0,1]\) by the prior range.
Red dots indicate the SW model estimates. Clearly, some parameters
change very little from permutation to permutation while others change
dramatically. \autoref{fig:perc-parameter-deviation} displays the same
parameter estimates as percent deviations from the SW model estimates.

\hypertarget{out-of-sample-forecasts}{%
\subsection{Out-of-sample forecasts}\label{out-of-sample-forecasts}}

\begin{figure}[t]

{\centering \includegraphics{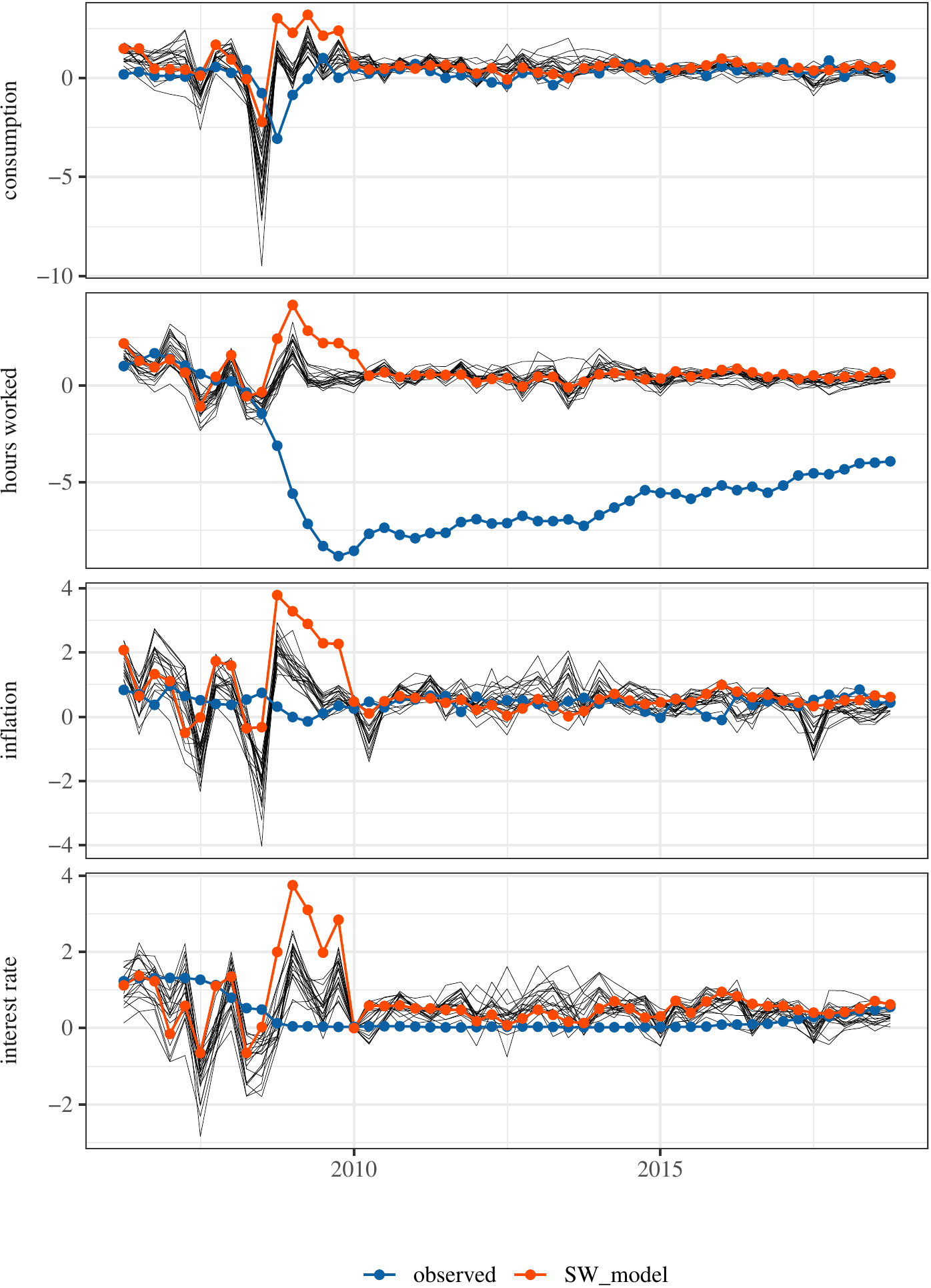} 

}

\caption{Out-of-sample predictions for the best predicting permutations (black), the SW model (red), and the observed data (blue).}\label{fig:pc-preds-p1}
\end{figure}

\begin{figure}[t]

{\centering \includegraphics{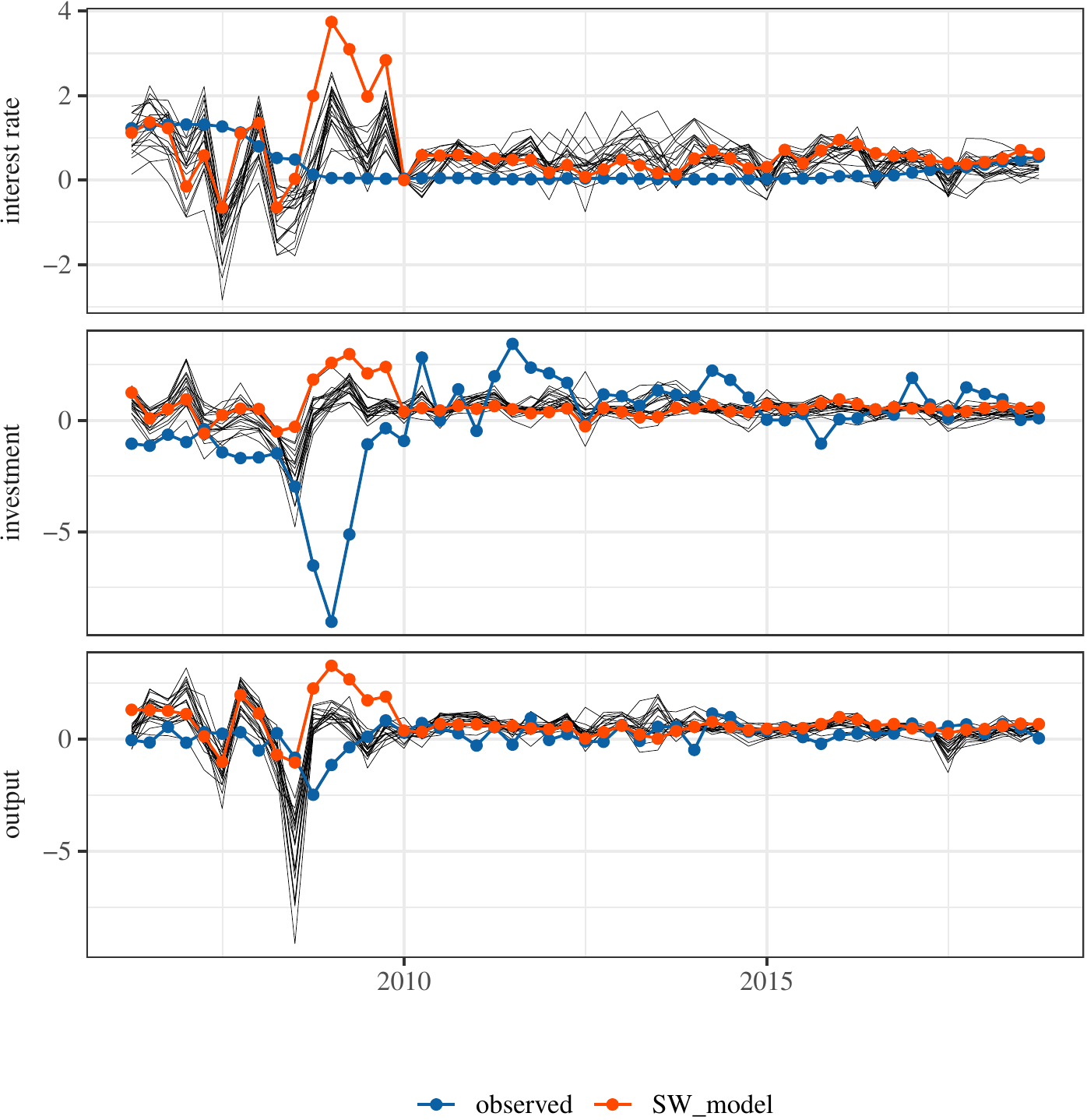} 

}

\caption{Out-of-sample predictions for the best predicting permutations (black), the SW model (red), and the observed data (blue).}\label{fig:pc-preds-p2}
\end{figure}

We now examine how well the SW model predicts future data relative to
the best models we could have used, had we seen the data.
\autoref{fig:pc-preds-p1} and \autoref{fig:pc-preds-p2} show the
out-of-sample predictions for top 20 flips based on ``average percent
improvement''. This analysis is a post-hoc measure as the best models
were selected to make these predictions well, though the parameters were
estimated without access to this data. We also show the observed data
and the predictions from the SW model. The SW model is quite bad at
predicting consumption, investment, output, and wages. It predicts a
consistent 0.5\% increase in the real wage level that never appears. In
fact, wages are far more volatile than any of the permutations can
account for. For the case of investment, output, and consumption, the SW
model drastically lags the 2009 recession and underestimates its
severity. Furthermore, it continues to predict a much stronger recovery,
even 10 years later, than has ever materialized. The SW model is quite
accurate for the interest rate, though this should perhaps be expected
given that the Taylor rule may well \emph{drive} Federal Reserve
decisions rather than \emph{describe} them.

\clearpage

\hypertarget{discussion}{%
\section{Discussion}\label{discussion}}

As we said in the introduction, there are very few who will defend the
forecasting record of DSGEs. Rather, their virtues are supposed to lie
in their capturing the structure of the economy, and so providing
theoretical insight, with meaningful parameters, and an ability to
evaluate policy and counterfactuals. We have examined these claims on
behalf of DSGEs through checking how well a DSGE can be estimated from
its own simulation output and by series permutation. In both cases, the
results are rather negative.

If we take our estimated model and simulate several centuries of data
from it, all in the stationary regime, and then re-estimate the model
from the simulation, the results are disturbing. Forecasting error
remains dismal and shrinks very slowly with the size of the data. Much
the same is true of parameter estimates, with the important exception
that many of the parameter estimates seem to be stuck around values
which differ from the ones used to generate the data. These ill-behaved
parameters include not just shock variances and autocorrelations, but
also the ``deep'' ones whose presence is supposed to distinguish a
micro-founded DSGE from mere time-series analysis or reduced-form
regressions. All this happens in simulations where the model
specification is correct, where the parameters are constant, and where
the estimation can make use of centuries of stationary data, far more
than will ever be available for the actual macroeconomy.

If we randomly re-label the macroeconomic time series and feed them into
the DSGE, the results are no more comforting. Much of the time we get a
model which predicts the (permuted) data \emph{better} than the model
predicts the unpermuted data. Even if one disdains forecasting as end in
itself, it is hard to see how this is at all compatible with a model
capturing something --- anything --- essential about the structure of
the economy.\footnote{It's conceivable that the issue is one of
  measurement. The variables in the DSGE model are defined by their
  roles in the inter-temporal optimization problem. (Thus labor makes a
  negative contribution to present utility but a positive contribution
  to present output, cannot be stored from period to period, etc.) The
  series gathered by the official statistical agencies use quite
  distinct definitions, and nothing guarantees that the series with
  analogous names are good measurements of the theoretical variables.
  (Cf. \citet[p. 6]{Haavelmo-probability-approach}: ``there is hardly an
  economist who feels really happy about identifying current series of
  `national income,' `consumption,', etc., with the variables by these
  names in his theories''.) Thus, perhaps, the model is right, but
  GDPDEF is really a better proxy for \emph{labor}, \(l_t\), than
  PRS85006023 is. This would, needless to say, raise its own set of
  difficulties for the interpretation and use of these models. One
  possible route forward would be to use \emph{many} observables which
  are all imperfectly aligned with the theoretical variables, which
  would, perhaps, require imposing a factor-model structure on the
  relations between observables and state variables.
  \citet{Boivin-Giannoni-DSGEs-in-data-rich-environment} is a step in
  this direction.} Perhaps even more disturbing, many of the parameters
of the model are essentially unchanged under permutation, including
``deep'' parameters supposedly representing tastes, technologies and
institutions.

To take this analysis one step further, we can examine the ``best''
predicting model, in terms of out-of-sample predictive log-likelihood.
Recall, this is just one of about 38\% which forecast better. It is also
important to note here, that the penalized negative log-likelihood
(in-sample) is very flat relative to permutations. In terms of that
metric, 19\% of permutations are within 10\% of the true permutation.
The true model arranges the data as

\begin{quotation}
hours worked, interest rate, inflation, output, consumption, investment, wages
\end{quotation}

\noindent while the best one by out-of-sample log-likelihood is

\begin{quotation}
hours worked, interest rate, output, wages, inflation, investment, consumption.
\end{quotation}

\noindent This model is pretty clearly scary to a macroeconomist. In the
Appendix, we take this model to most implausible extreme. We rewrite the
introduction of \citet{SmetsWouters2007}, permuting all the series to
reflect the best one by negative predictive log likelihood. Such a
description of the macroeconomy is doubtless complete nonsense, and yet
this interpretation would have led to better predictions of
macroeconomic comovements.

Ignoring the fact that the DSGE provides generally poor forecasts
relative to alternative, yet uninterpretable, similar models, one might
wonder whether it manages to avoid the Lucas Critique. That is, can
policy makers imagine that the model truly decouples policy variables
from the ``deep'' parameters that economic actors maintain?
Specifically, for example, if the Fed changes how they manage interest
rates, do the other parameters move? This question can be investigated
directly by examining the distribution of Taylor Rule parameters
conditional on the truth generating the model and estimating said
parameters in the simulation exercise.\footnote{This analysis is not
  constrained to the distribution of plausible Taylor rule parameters
  that the Fed might consider, but rather simply investigates all
  possible such parameters.} These correlations are shown in
\autoref{tab:covary-with-Taylor}. The Taylor rule parameters are,
\(\rho\), the autocorrelation in the interest rate, \(r_\pi\), the
response to inflation, \(r_y\), the response to deviations of output
from potential, and \(r_{\Delta y}\) the response to changes in the
deviation. A handful of deep parameters have correlations larger in
magnitude than 0.3 (shown in bold). This magnitude may or may not be
large enough for concern, but the correlation is also not negligible. So
the question of whether these models actually address their original
motivation is, at the very least, still open.

\begin{table}

\caption{\label{tab:covary-with-Taylor}Correlations between ``deep'' parameters and Taylor rule parameters from the ``Simulate and estimate'' exercise.}
\centering
\begin{tabular}[t]{rrrrr}
\toprule
 & $r_\pi$ & $\rho$ & $r_y$ & $r_{\Delta y}$\\
\midrule
$\phi_1$ & 0.2 & 0.1 & -0.03 & 0.11\\
$\sigma_c$ & 0.06 & 0.26 & 0.17 & 0.02\\
$h$ & \textbf{0.33} & 0.08 & 0.02 & \textbf{-0.34}\\
$\xi_w$ & -0.06 & \textbf{0.43} & 0.02 & 0.08\\
$\sigma_l$ & 0.01 & 0.19 & 0.03 & \textbf{-0.37}\\
\addlinespace
$\xi_p$ & -0.26 & \textbf{0.36} & 0.13 & 0.19\\
$\iota_w$ & 0.1 & 0.04 & 0.05 & 0.01\\
$\iota_p$ & 0.03 & 0.17 & -0.03 & 0.04\\
$\Psi$ & \textbf{0.34} & 0.01 & 0.1 & \textbf{0.33}\\
$\Phi$ & 0.02 & 0.03 & 0.03 & 0.02\\
\addlinespace
$r_\pi$ & \textbf{1} & \textbf{0.31} & 0.24 & 0\\
$\rho$ & \textbf{0.31} & \textbf{1} & \textbf{0.32} & 0.12\\
$r_y$ & 0.24 & \textbf{0.32} & \textbf{1} & -0.04\\
$r_{\Delta y}$ & 0 & 0.12 & -0.04 & \textbf{1}\\
$\overline{\pi}$ & -0.04 & -0.05 & -0.08 & 0.01\\
\addlinespace
$100(\beta^{-1} -1)$ & -0.04 & -0.18 & -0.14 & -0.04\\
$\overline{l}$ & 0.07 & 0.1 & 0.08 & -0.05\\
$\overline{\gamma}$ & 0.02 & 0.03 & 0.05 & -0.01\\
$\rho_{ga}$ & -0.23 & -0.15 & 0 & 0.01\\
$\alpha$ & 0.18 & 0.11 & 0.16 & -0.04\\
\bottomrule
\end{tabular}
\end{table}

The results of the two tests described here are grim. Series swapping
gives us strong reasons to doubt that the DSGE machinery manages to
capture anything important about the structure of the economy. Even if
one dismisses that, and believes (perhaps because ``theory is evidence
too'') that the DSGE must be right, the simulation exercise shows that,
even under the most favorable possible circumstances, it is simply wrong
to think that the DSGE will give reasonably accurate predictions, or
even that it can be reliably estimated.

We have not, of course, proved that flaws like this are inherent in the
DSGE form. But we have \emph{not} cherry-picked an obsolete or marginal
model.\footnote{We obtained very similar results for the real-business-cycle
model of \citet{KydlandPrescott1982}, but omit them here, because that model
\emph{is} obsolete.} The SW model is widely regarded as the baseline
DSGE for the economy of the United States, which is by far the most
important national economy in the world. The SW paper has been cited
over 6300 times.\footnote{Google Scholar
(\url{https://scholar.google.com/scholar?cites=8854430771281116653}), accessed
27 October 2022.} What concerns us is that in all that literature, we
appear to be the first to have subjected it to such direct, even
elementary, tests. We do not assert that all DSGE models must be
pathological in the ways we have shown the SW model is. Indeed, readers
are free to hope that their favorite DSGE does capture economic
structure and can be meaningfully estimated with reasonable amounts of
data. But we hope we have persuaded readers that they can, and should,
do more than hope: they can \emph{check}.

\clearpage

\appendix

\hypertarget{sec:data-preprocessing}{%
\section{Data preprocessing}\label{sec:data-preprocessing}}

\footnotesize

The necessary series are shown in \autoref{tab:data}. All of the data
are quarterly. The required series are GDPC1, GDPDEF, PCEC, FPI, CE16OV,
FEDFUNDS, CNP16OV, PRS85006023, and COMPNFB. These nine series are used
to create \(x_t\) as follows: \begin{align*}
  pop_t^{ind} &= CNP16OV_t / CNP16OV_{2012Q3},\\
  emp_t^{ind} &= CE16OV_t / CE16OV_{2012Q3},\\
  y_t &= 100\ln\left(\frac{GDPC1_t}{pop_t^{ind}}\right),\\
  c_t &= 100\ln\left(\frac{PCEC_t/GDPDEF_t}{pop_t^{ind}}\right),\\
  i_t &= 100\ln\left(\frac{FPI_t/GDPDEF_t}{pop_t^{ind}}\right),\\
  l_t &=
        100\ln\left(\frac{PRS85006023_t/emp_t^{ind}}{pop_t^{ind}}\right),\\
  \pi_t &= 100\ln\left(\frac{GDPDEF_t}{GDPDEF_{t-1}}\right),\\
  w_t &= 100\ln\left(\frac{COMPNFB_t}{GDPDEF_t}\right),\\
  r_t &= FEDFUNDS/4.
\end{align*}

\begin{table}[h]
  \centering
  \resizebox{\linewidth}{!}{
  \begin{tabular}{@{}lll@{}}
    \toprule
    Series ID & Description & Unit \\
    \midrule
    GDPC1 & Real Gross Domestic Product & Billions of Chained 2012 \$\\
    GDPDEF & GDP Implicit Price Deflator & Index: 2012=100\\
    PCEC & Personal Consumption Expenditures & Billions of \$\\
    FPI & Fixed Private Investment & Billions of \$\\
    CE16OV & Civilian Employment & Thousands of persons\\
    FEDFUNDS & Effective Federal Funds Rate & Percent\\
    CNP16OV & Civilian Noninstitutional Population & Thousands of
                                                     persons\\
    PRS85006023 & Nonfarm business sector: average weekly hours &
                                                                  Index:
                                                                  2012=100\\
    COMPNFB & Nonfarm business sector: Compensation per hour & Index: 2012=100\\
    \bottomrule
  \end{tabular}
  }
  \caption{Data series from FRED for estimating the DSGE.}
  \label{tab:data}
\end{table}

\clearpage

\hypertarget{rewriting}{%
\section{\texorpdfstring{Rewriting
\citet{SmetsWouters2007}}{Rewriting }}\label{rewriting}}

In the rest of this section, we describe the log-linearized version of
the DSGE model that we subsequently estimate using US data. All
variables are log-linearized around their steady-state balanced growth
path. Starred variables denote steady-state values. We first describe
the aggregate demand side of the model and then turn to the aggregate
supply.

The aggregate resource constraint is given by \begin{equation}
\label{eq:1}
\yobs{t} =   c_y\cobs{t} +  i_y \iobs{t} + z_y z_t +
\epsilon^g_t.
\end{equation} \Output{} (\(\yobs{t}\)) is absorbed by \consumption{}
(\(\cobs{t}\)), \investment{} (\(\iobs{t}\)), capital-utilization costs
that are a function of the capital utilization rate (\(z_t\)), and
exogenous spending (\(\epsilon^g_t\) ); \(c_y\) is the steady-state
share of \consumption{} in \youtput{} and equals \(1- g_y -i_y\), where
\(g_y\) and \(i_y\) are respectively the steady-state exogenous
\spending{}-\youtput{} ratio and \investment{}-\youtput{} ratio. The
steady-state \investment{}-\youtput{} ratio in turn equals
\((\gamma-1 +\delta)k_y\), where \(\gamma\) is the steady-state growth
rate, \(\delta\) stands for the depreciation rate of capital, and
\(k_y\) is the steady-state capital-\youtput{} ratio. Finally,
\(z_y= R_*^k ky\), where \(R_*^k\) is the steady-state rental rate of
capital. We assume that exogenous \spending{} follows a first-order
autoregressive process with an IID-Normal error term and is also
affected by the productivity shock as fol- lows:
\(\epsilon^g_t = \rho_g\epsilon^g_{t-1} + \eta_t^g +\rho_{ga}\eta_t^a\).
The latter is empirically motivated by the fact that, in estimation,
exogenous \spending also includes net exports, which may be affected by
domestic productivity developments.

The dynamics of \consumption{} follows from the \consumption{} Euler
equation and is given by \begin{equation}
  \cobs{t} = c_1\cobs{t-1} + (1-c_1)\E_t[\cobs{t+1}] + c_2 (\lobs{t}
  -\E_t[\lobs{t+1}]) - c_3(\robs{t}-\E_t[\piobs{t+1}])+\epsilon_t^b),
\end{equation} where \(c_1 = (\lambda/\gamma)/(1+\lambda/\gamma)\),
\(c_2 = [(\sigma_c - 1)(W_*^h L_*/C_*)] / [\sigma_c(1+\lambda/\gamma)],\)
and \(c_3 = (1-\lambda/\gamma)/[\sigma_c(1+\lambda/\gamma)]\). Current
\consumption{} \((\cobs{t}\) depends on a weighted average of past and
expected future \consumption, and on expected growth in \hours{}
\(\lobs{t}  -\E_t[\lobs{t+1}])\), the ex ante \realinterest{}
\((\robs{t}-\E_t[\piobs{t+1}]\), and a disturbance term \(\epsilon^b_t\)
. Under the assumption of no external habit formation \((\lambda=0)\)
and log utility in \consumption{} \((\sigma_c =1)\), \(c1 = c2 = 0\) and
the traditional purely forward-looking \consumption{} equation is
obtained. With steady-state growth, the growth rate \(\gamma\)
marginally affects the reduced-form parameters in the linearized
\consumption{} equation. When the elasticity of intertemporal
substitution (for constant \labor) is smaller than one
\((\sigma_c > 1)\), \consumption{} and \hours{} are complements in
utility and \consumption{} depends positively on current \hours{} and
negatively on expected growth in \hours{} (see Susanto Basu and Kimball
2002). Finally,the disturbance term \(\epsilon_t^b\) represents a wedge
between the \interest{} controlled by the central bank and the return on
assets held by the households. A positive shock to this wedge increases
the required return on assets and reduces current \consumption. At the
same time, it also increases the cost of capital and reduces the value
of capital and \investment, as shown below.

The dynamics of \investment{} comes from the \investment{} Euler
equation and is given by \begin{equation}
  \iobs{t} = i_1\iobs{t-1} + (1-i_1)\E[\iobs{t+1}] + i_2 q_t + \epsilon_t^i,
\end{equation} where \(i_1 = 1/(1+\beta\gamma^{1-\sigma_c}),\)
\(i_2 = 1/[(1+\beta\gamma^{1-\sigma_c})\gamma^2\varphi]\), \(\varphi\)
is the steady-state elasticity of the capital adjustment cost function,
and is the discount factor applied by households. As in
\citet{ChristianoEichenbaum2005}, a higher elasticity of the cost of
adjusting capital reduces the sensitivity of \investment{}
(\(\iobs{t}\)) to the real value of the existing capital stock
(\(q_t\)). Modeling capital adjustment costs as a function of the change
in \investment{} rather than its level introduces additional dynamics in
the \investment{} equation, which is useful in capturing the hump-shaped
response of \investment{} to various shocks. Finally, \(\epsilon^i_t\)
represents a disturbance to the \investment-specific technology process
and is assumed to follow a first-order autoregressive process with an
IID-Normal error term:
\(\epsilon_t^i = \rho_i\epsilon_{t-1}^i + \eta_t^i\).

The corresponding arbitrage equation for the value of capital is given
by \begin{equation}
q_t = q_1\E_t[q_{t+1}] + (1-q_1)\E_t[r_{t+1}^k] - (\robs{t} -
\E[\piobs{t+1}] + \epsilon_t^b),
\end{equation} where
\(q_1 = \beta\gamma^{-\sigma_c}(1-\delta) = [(1-\delta)/(R_*^k + (1-\delta))]\).
The current value of the capital stock (\(q_t\)) depends positively on
its expected future value and the expected real rental rate on capital
(\(\E_t[r_{t+1}^k]\)) and negatively on the ex ante \realinterest{} and
the risk premium disturbance.

Turning to the supply side, the aggregate production function is given
by \begin{equation}
  \yobs{t} = \phi_p(\alpha k_t^2 + (1-\alpha)\lobs{t} + \epsilon_t^a).
\end{equation} \Output{} is produced using capital (\(k_t^s\)) and
\labor{} services (\hours, \(\lobs{t}\)). Total factor productivity
(\(\epsilon_t^a\)) is assumed to follow a first-order autoregressive
process: \(\epsilon_t^a = \rho_a \epsilon_{t-1}^a + \eta_t^a\). The
parameter \(\alpha\) captures the share of capital in production, and
the parameter \(\phi_p\) is one plus the share of fixed costs in
production, reflection the presence of fixed costs in production.

As newly installed capital becomes effective only with a one-quarter
lag, current capital services used in production \((k_t^s)\) are a
function of capital installed in the previous period \((k_{t-1})\) and
the degree of capital utilization \((z_t)\): \begin{equation}
  k_t^s = k_{t-1} + z_t.
\end{equation} Cost minimization by the households that provide capital
services implies that the degree of capital utilization is a positive
function of the rental rate of capital, \begin{equation}
  \label{eq:7}
  z_t = z_1 r_t^k,
\end{equation} where \(z_q = (1-\psi)/\psi\) and \(\psi\) is a positive
function of the elasticity of the capital utilization adjustment cost
function and normalized to be between zero and one. When \(\psi=1\), it
is extremely costly to change the utilization of capital and, as a
result, the utilization of capital remains constant. In contrast, when
\(\psi=0\), the marginal cost of changing the utilization of capital is
constant and, as a result, in equilibrium the rental rate on capital is
constant, as is clear from equation \eqref{eq:7}.

The accumulation of installed capital \((k_t)\) is a function not only
of the flow of \investment{} but also of the relative efficiency of
these \investment{} expenditures as captured by the \investment-specific
technology disturbance \begin{equation}
  \label{eq:8}
  k_t = k_1 k_{t-1} + (1-k_1) \iobs{t} + k_2 \epsilon_{t}^i,
\end{equation} with \(k_1=(1-\delta)/\gamma\) and
\(k_2 = (1-(1-\delta)/\gamma) (1+ \beta\gamma^{1-\sigma_c})\gamma^2\varphi.\)

Turning to the monopolistic competitive goods market, cost minimization
by firms implies that the price mark-up (\(\mu_t^p\)), defined as the
difference between the average price and the nominal marginal cost or
the negative of the real marginal cost, is equal to the difference
between the marginal product of \labor{} (\(mp\lobs{t}\)) and the real
\wage{} (\(\wobs{t}\)): \begin{equation}
  \label{eq:9}
  \mu_t^p = mp\lobs{t} - \wobs{t} = \alpha(k_t^s - \lobs{t}) +
  \epsilon_t^a - \wobs{t}.
\end{equation} As implied by the second equality in \eqref{eq:9}, the
marginal product of \labor{} is itself a positive function of the
capital-\labor{} ratio and total factor productivity.

Due to \price{} stickiness, as in Calvo (1983), and partial indexation
to lagged \inflation{} of those \prices{} that can not be reoptimized,
as in Smets and Wouters (2003), \prices{} adjust only sluggishly to
their desired mark-up. Profit maximization by \price-setting firms gives
rise to the following New-Keynesian Phillips curve: \begin{equation}
  \label{eq:10}
  \piobs{t} = \pi_1\piobs{t-1} + \pi_2 \E_t[\piobs{t+1}] - \pi_3
  \mu_t^p + \epsilon_t^p,
\end{equation} where
\(\pi_1 = \iota_p / (1+ \beta\gamma^{1-\sigma_c} \iota_p),\)
\(\pi_2 = \beta \gamma^{1-\sigma_c} / (1+\beta \gamma^{1-\sigma_c} \iota_p),\)
and
\(\pi_3 = 1/(1 +\beta \gamma^{1-\sigma_c} \iota_p)[(1- \beta \gamma^{1-\sigma_c} \xi_p)(1-\xi_p) / \xi_p ((\phi_p - 1)\epsilon_p +1)].\)
\Inflation{} \((\piobs{t})\) depends positively on past and expected
future \inflation, negatively on the current \price{} mark-up, and
positively on a \price{} mark-up disturbance \((\epsilon_t^p)\). The
\price{} mark-up disturbance is assumed to follow an ARMA(1,1) process:
\(\epsilon_t^p = \rho_p \epsilon_{t-1}^p + \eta_t^p - \mu_p \eta_{t-1}^p\),
where \(\eta_t^p\) is an IID-Normal \price mark-up shock. The inclusion
of the MA term is designed to capture the high-frequency fluctuations in
\inflation.

When the degree of indexation to past \inflation{} is zero
(\(\iota_p =0\)), equation \eqref{eq:10} reverts to a standard, purely
forward-looking Phillips curve (\(\pi_1 = 0\)). The assumption that all
\prices{} are indexed to either lagged \inflation{} or the steady-state
\inflationrate{} ensures that the Phillips curve is vertical in the long
run. The speed of adjustment to the desired mark-up depends, among
others, on the degree of \price-stickiness (\(\xi_p\)), the curvature of
the Kimball goods market aggregator (\(\epsilon_p\)), and the
steady-state mark-up, which in equilibrium is itself related to the
share of fixed costs in production (\(\phi_p - 1\)) through a
zero-profit condition. A higher \(\epsilon_p\) slows down the speed of
adjustment because it increases the strategic complementarity with other
\price{} setters. When all \prices{} are flexible (\(\xi_p = 0\)) and
the \price-mark-up shock is zero, equation \eqref{eq:10} reduces to the
familiar condition that the \price{} mark-up is constant, or
equivalently that there are no fluctuations in the wedge between the
marginal product of \labor{} and the real \wage.

Cost minimization by firms will also imply that the rental rate of
capital is negatively related to the capital-\labor{} ratio and
positively to the real \wage{} (both with unitary elasticity):
\begin{equation}
  \label{eq:11}
  r_t^k = -(k_t - \lobs{t}) + \wobs{t}
\end{equation} In analogy with the goods market, in the monopolistically
competitive \labor{} market, the \wage{} mark-up will be equal to the
difference between the real \wage{} and the marginal rate of
substitution between \working{} and \consuming{} (\(mrs_t\)),
\begin{equation}
  \label{eq:12}
  \mu_t^w = \wobs{t} - mrs_t = \wobs{t} - (\sigma_l\lobs{t} +
  \frac{1}{1-\lambda/\gamma} \left(\cobs{t} - \frac{\lambda\cobs{t-1}}{\gamma}\right),
\end{equation} where \(\sigma_l\) is the elasticity of \labor{} supply
with respect to the real \wage{} and \(\lambda\) is the habit parameter
in \consumption.

Similarly, due to nominal \wage{} stickiness and partial indexation of
\wages{} to \inflation, real \wages{} adjust only gradually to the
desired \wage{} mark-up: \begin{equation}
  \label{eq:13}
  \wobs{t} = w_1\wobs{t-1} + (1- w_1)(\E_t[\wobs{t+1}] +
  \E_t[\piobs{t+1}]) - w_2 \piobs{t} + w_3 \piobs{t-1} - w_4 \mu_t^w + \epsilon_t^w,
\end{equation} with \(w_1 = 1/(1+\beta\gamma^{1-\sigma_c})\),
\(w_2= (1+\beta\gamma^{1-\sigma_c}\iota_w) / (1+\beta\gamma^{1-\sigma_c})\),
\(w_3 = \iota_w / (1+\beta\gamma^{1-\sigma_c}),\) and
\(w_4 = 1/ (1+\beta\gamma^{1-\sigma_c}) [(1-\beta\gamma^{1-\sigma_c}\xi_w)(1-\xi_w) / (\xi_w((\phi_w-1)\epsilon_w + 1))]\).

The real \wage{} \(\wobs{t}\) is a function of expected and past real
\wages, expected, current, and past \inflation, the \wage{} mark-up, and
a \wage-markup disturbance \((\epsilon^w_t)\). If \wages{} are perfectly
flexible \((\xi_w = 0)\), the real \wage{} is a constant mark-up over
the marginal rate of substitution between \consumption{} and \leisure{}.
In general, the speed of adjustment to the desired \wage{} mark-up
depends on the degree of \wage{} stickiness \((\xi_w)\) and the demand
elasticity for \labor, which itself is a function of the steady-state
\labor{} market mark-up \((\phi_w - 1)\) and the curvature of the
Kimball \labor{} market aggregator \((\epsilon_w)\). When \wage{}
indexation is zero \((\iota_w =0)\), real \wages{} do not depend on
lagged \inflation{} \((w_3 = 0)\). The \wage{}-markup disturbance
\((\epsilon_w^t)\) is assumed to follow an ARMA(1, 1) process with an
IID-Normal error term:
\(\epsilon_t^w = \rho_w \epsilon_{t-1}^w + \eta_t^w - \mu_w \eta_{t-1}^w\).
As is the case of the \price{} mark-up shock, the inclusion of an MA
term allows us to pick up some of the high-frequency fluctuations in
\wages{}.\footnote{Alternatively, we
  could interpret this disturbance as a \labor{} supply disturbance
  coming changes in preferences for \leisure{}.}

Finally, the model is closed by adding the following empirical monetary
policy reaction function: \begin{align}
  \label{eq:14}
  \robs{t} &= \rho\robs{t-1} + (1-\rho)\left[r_\pi \piobs{t} +
  r_y(\yobs{t}-\yobs{t}^p)\right]+\\ &\quad + r_{\Delta y}\left[(\yobs{t}-\yobs{t}^{p}) -
  (\yobs{t-1} - \yobs{t-1}^{p})\right] + \epsilon_t^r.\notag
\end{align}

The monetary authorities follow a generalized Taylor rule by gradually
adjusting the policy-controlled \interest{} \((\robs{t})\) in response
to \inflation{} and the \youtput{} gap, defined as the difference
between actual and potential \youtput{} \citep{Taylor-on-Taylor-rule}.
Consistently with the DSGE model, potential \youtput{} is defined as the
level of \youtput{} that would prevail under flexible \prices{} and
\wages{} in the absence of the two ``mark-up'\,'
shocks.\footnote{In practical
  terms, we expand the model consisting of equations \eqref{eq:1} to
  \eqref{eq:14} with a flexible-\price-and-\wage{} version in order to
  calculate the model-consistent \youtput{} gap. Note that the assumption
  of treating the \wage{} equation disturbance as a \wage{} mark-up
  disturbance rather than a \labor{} supply disturbance coming from
  changed preferences has implications for our calculation of
  potential \youtput{}. }

The parameter \(\rho\) captures the degree of \interest{} smoothing. In
addition, there is a short-run feedback from the change in the
\youtput{} gap. Finally, we assume that the monetary policy shocks
\((\epsilon_t^r)\) follow a first-order autoregressive process with an
IID-Normal error term:
\(\epsilon_t^r = \rho_r\epsilon_{t-1}^r + \eta_t^r\).

Equations \eqref{eq:1} to \eqref{eq:14} determine 14 endogenous
variables: \(y_t,\ c_t,\ i_t,\) \(q_t,\ k_t^s,\) \(k_t,\ z_t,\ r_t^k,\)
\(\mu_t^p,\ \pi_t,\ \mu_t^w,\ w_t,\ l_t,\) and \(r_t\). The stochastic
behavior of the system of linear rational expectations equations is
driven by seven exogenous disturbances: total factor productivity
\((\epsilon^a_t)\), investment-specific technology \((\epsilon^i_t)\),
risk premium \((\epsilon_t^b)\), exogenous spending \((\epsilon_t^g)\),
\price{} mark-up \((\epsilon_t^p)\), \wage{} mark-up \((\epsilon_t^w)\),
and monetary policy \((\epsilon_t^r)\) shocks. Next we turn to the
estimation of the model.

\bibliographystyle{plainnat}
\bibliography{dsges.bib}

\end{document}